\documentclass{article}
\usepackage[utf8]{inputenc}

\usepackage{amsmath}
\usepackage{amssymb}
\usepackage{subcaption}
\usepackage{mathtools}
\usepackage{graphicx}
\usepackage{color}
\usepackage{authblk}
\usepackage[urlcolor=blue,bookmarks=false,hypertexnames=true]{hyperref}
\usepackage{geometry}
\usepackage{float}
\usepackage[titletoc,title]{appendix}
\usepackage{cancel}
\usepackage{tabularx}
\usepackage{colortbl}
\usepackage{dcolumn}
\usepackage{bm}
\usepackage{latexsym}
\usepackage{psfrag}
\usepackage{epsfig}
\usepackage{epstopdf}
\usepackage{psfrag}
\usepackage{tikz} 
\usepackage{color}
\usepackage{braket}
\usepackage[english]{babel}
\usepackage{array}
\usepackage{authblk}
\usepackage{blindtext}
\usepackage{titling}

\title{Test of Quantum Gravity in Statistical Mechanics}

\date{}

\begin{document}

\author{Saurya Das \thanks{saurya.das@uleth.ca}\hspace{0.1cm}} 

\author{Mitja Fridman \thanks{fridmanm@uleth.ca}}
\affil{Theoretical Physics Group and Quantum Alberta, \\
Department of Physics and Astronomy,
University of Lethbridge, \\
4401 University Drive, Lethbridge,
Alberta, T1K 3M4, Canada}

\maketitle

\begin{abstract}
    We study Quantum Gravity effects on the density of states in statistical mechanics and its implications for the critical temperature of a Bose Einstein Condensate and fraction of bosons in its ground state. 
    We also study the effects of compact extra dimensions on the critical temperature and the fraction. We consider both neutral and charged bosons in the study and show that the effects may just be measurable in current and future experiments. 

    %
\end{abstract}

\tableofcontents
\newpage

\section{Introduction}

Quantum Mechanics (QM) and General Relativity (GR) successfully describe observable phenomena in microscopic and macroscopic regimes respectively. 
%
%
However, there is no simple way of combining these two theories to describe 
phenomena in regimes where both theories are applicable. 
Candidate theories of Quantum Gravity (QG), which aim to accomplish this, such as
String Theory and Loop Quantum Gravity, have made significant progress. Yet, there has not been a single experiment or observation which support or refute any QG theory. 
Since the immensity of the QG or the `Planck scale', about $10^{16}\,\mathrm{TeV}$, prevents such tests directly in colliders, it is important to look for indirect signatures of these theories in accessible, low energy laboratory based experiments \cite{DV,GAC}. Potential QG signatures in condensed matter, atomic and molecular experiments have been explored by various authors \cite{CM1,CM2,ADV0,CM3,CM4,CM5,CM6,CM7}. However, QG effects in statistical mechanical systems have not been studied extensively,
although there has been some work 
%
done in the  search for QG signatures in BECs \cite{BGL,FB,RCG}
and that of compactified extra dimensions on BECs \cite{shiraishi,AA,LB}. 
We also study QG effects in BEC in this paper, although our approach is different and we obtain a number of new results.  
%
%
%
%
Furthermore, we will study the feature that most QG theories predict, namely a minimum measurable length and the related Generalized Uncertainty Principle (GUP) \cite{ADV0,GUP1,GUP2,GUP3,GUP4,KMM,FC,IPK,SLV,KP}. 
We will work with the general form of the GUP, incorporating both linear and quadratic terms in momenta, and which imply a minimum measurable length and a maximum measurable momentum, is given by \cite{ADV0}
 \begin{equation}
 \label{gup}
     [x_i,p_j]=i\hbar\left(\delta_{ij}-\alpha\left(p\delta_{ij}+\frac{p_ip_j}{p}\right)+\beta\left(p^2\delta_{ij}+3p_ip_j\right)\right)~.
 \end{equation}
In the above, $x_i$ and $p_j$ are the position and momentum operators respectively and 
 $p=\sqrt{p_ip_i}$. Also, $\alpha \equiv \alpha_0/(M_{P}c)$, 
 $\beta \equiv \beta_0/(M_{P}c)^2$, 
where 
 $\alpha_0$ and $\beta_0$ are the linear and quadratic GUP parameters 
 %
 and $M_{P}=\sqrt{\hbar c/G}$ is the Planck mass.
 It is sometimes assumed that $\alpha_0,\beta_0 = {\cal O}(1)$. However, we will
 make no such restriction and compute QG effects for arbitrary 
 $\alpha_0$ and $\beta_0$. Note that this implies intermediate length scales 
 $\alpha_0\, \ell_{Pl}$ and  $\sqrt{\beta_0}\, \ell_{Pl}$ between the electroweak length scale ($\approx 10^{-18}$ m) and the Planck scale, $\ell_{Pl} \approx 10^{-35}$ m. The only restrictions that these impose are
 $\alpha_0< 10^{17}$ and $\beta_0 < 10^{34}$, the bounds implied indirectly by LHC experiments, since no new fundamental length (or energy) scale has been observed therein.
 
The above, when applied to Statistical Mechanics, modifies the energy levels of a particle in a box, and hence the corresponding
phase space volume of a quantum particle in a box \cite{pathria}.
This when applied to the statistics of a 
BEC, modifies its critical temperature $T_c$ and the fraction of bosons in the ground state at any $0<T<T_c$. 
%
Using the above, in this paper we present a new approach to computing QG corrections to observables in BEC, such as the critical temperature and fraction of bosons in the ground state. This approach modifies the density of states by approximating it to first order in $\alpha$ and $\beta$. %

This paper is organized as follows. In section \ref{SBEC}, we present the standard results of the BEC, which is followed by a discussion about the role of compact dimensions in BEC in section \ref{BECcd}. QG effects on observables in BEC and their potential measurability 
are discussed in section \ref{QGBEC}. 
We summarize the work in section \ref{conc}.
 
 \section{Standard Bose-Einstein condensation}
 \label{SBEC}

The phenomenon of BEC occurs when a dilute gas of bosons is cooled below a certain temperature, 
such that more and more bosons start occupying the ground state. This temperature is known as the critical temperature $T_c$. Since Bose-Einstein (BE) statistics (see Eq.(\ref{befdd}) in Appendix \ref{mtsm}) allows for an arbitrary number of bosons in any state, there could theoretically be an infinite number of bosons in the ground state. 
We review a few important results related to a BEC that will be used in the rest of the paper. Note that results in this section for $T_c$ are valid for arbitrary $d$ spatial dimensions.
%
%

The critical temperature $T_c$ is the threshold at which one still has all the bosons in the excited states.
As the gas temperature $T$ is decreased from $T_c$, they start dropping to the ground state. 
Furthermore, the chemical potential 
$\mu\rightarrow0$ at $T=T_c$
in the non-relativistic case, which we consider first.
%
In $d$-dimensional space, with $d\geq 3$
(since it can be shown that there can be no non-relativistic BEC in 1 and 2 dimensions \cite{AA}), 
the critical temperature takes the form
\begin{eqnarray}
\label{edtc}
    T_c=\frac{2\pi\hbar^2}{k_Bm\zeta(\tfrac{d}{2})^{2/d}}n^{2/d}~,
\end{eqnarray}
%
from which one can see that the critical temperature of a non-relativistic BEC will be higher for high boson densities and light boson masses. 
%
%
The second important observable in BEC is the fraction 
$f_0$ of bosons in the ground state.
If $n_0$ is the number density of bosons in the ground state, $n(T)$ the number density in the excited states at temperature $T<T_c$ and $n$ the total number density, 
then these are related by
%
%
%
\begin{equation}
\label{fderiv}
    n=n_0+n(T)=n_0+n\left(\frac{T}{T_c}\right)^{d/2}\,\,\implies\,\,\,\,f_0=\frac{n_0}{n}=1-\left(\frac{T}{T_c}\right)^{d/2}~.
\end{equation}
%
From Eq.(\ref{fderiv}) we can see, that at $T=T_c$, all bosons are still in the excited states, since $f_0=0$. The bosons start to occupy the ground state for $T<T_c$, when $f_0>0$ and completely fill the ground state at $T=0\,\mathrm{K}$, when $f_0=1$.

The critical temperature and fraction of bosons in the ground state for the relativistic case can be found in a similar manner. We consider two cases of relativistic bosons. The first, when they are considered neutral, is 
associated with the following critical temperature
%
%
\begin{eqnarray}
\label{rbtc}
T_c=\frac{1}{k_B}\left(\frac{2^{d-1}\,\pi^{d/2}\,\hbar^d\,c^d\,\Gamma(\tfrac{d}{2})}{\Gamma(d)\,\zeta(d)}\right)^{1/d}n^{1/d}
\end{eqnarray}
for arbitrary $d\geq2$ spatial dimensions \cite{GLB,Pandita}.
%
Note that while it is dependent on the number density of bosons, it does not depend on boson mass, unlike the non-relativistic result
given in Eq.(\ref{edtc}). 
However, it continues to depend on the boson number density, albeit with a different (positive) power. 
%
%
In this case, the fraction of relativistic neutral bosons in the ground state turns out to be
\begin{eqnarray}
\label{rbfr}
f_0=\frac{n_0}{n}=1-\left(\frac{T}{T_c}\right)^d.
\end{eqnarray}
The second relativistic case includes both bosons and antibosons. 
The distribution function for this case is obtained by subtracting two BE distributions, 
one for bosons $\mu(T_c)=mc^2$ and one for antibosons $\mu(T_c)=-mc^2$, to compute the total charge density $n$ (in previous cases this was just the number density). The relativistic boson-antiboson critical temperature can be expressed in arbitrary dimensional Euclidean space $d\geq 3$
as \cite{GLB,Pandita}
\begin{eqnarray}
\label{rbbtc}
T_c=\frac{1}{k_B}\left(\frac{2^{d-2}\,\pi^{d/2}\,\hbar^d\,c^{d-2}\,\Gamma(\tfrac{d}{2})}{m\,\Gamma(d)\,\zeta(d-1)}\right)^{1/(d-1)}n^{1/(d-1)}~.
\end{eqnarray}
%
Note that the above depends on the boson mass,
and the number density. The critical temperature 
increases with increasing number density and decreasing boson mass. 
%
The fraction of bosons in the ground state in this case
turns out to be
\begin{eqnarray}
\label{rbbfr}
f_0=\frac{n_0}{n}=1-\left(\frac{T}{T_c}\right)^{(d-1)}.
\end{eqnarray}
Note the different power of $T/T_c$ when compared with 
Eq.(\ref{rbfr}).

To summarize, we have seen in this section that the BEC
critical temperature in all cases is 
a function of boson mass $m$ and boson number density $n$, 
with different powers for different cases. 
Similarly, the fraction of bosons in the ground state 
depends on different powers of $T/T_c$ for the different cases.


\section{Bose-Einstein condensation in compact dimensions}
\label{BECcd}

\begin{figure}[h]
     \centering
     \includegraphics[scale=0.95]{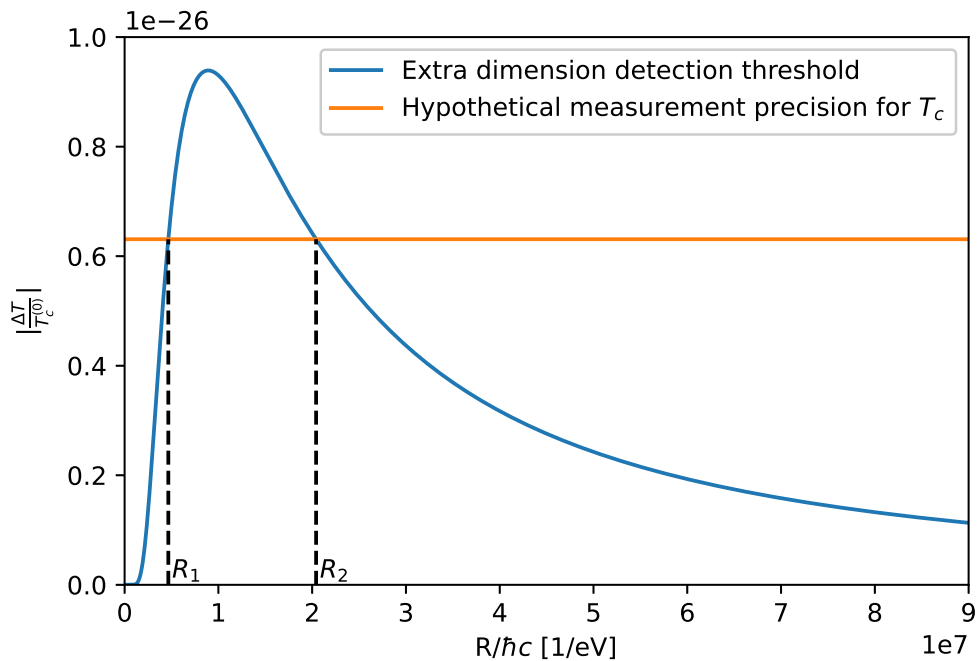}
     \caption{Relative correction as a function of radius $R$ of the compact dimension for a helium gas, using $m=m_{He}$, $n=10^{23}\,\mathrm{m}^{-3},$ $d=3$ and $N=1$ (blue), and a hypothetical precision of the experiment (orange).}
     \label{cdfig}
 \end{figure}

Compact extra dimensions are interesting from the point of view of QG, since they are an essential component in String Theory, where they are normally assumed to be tiny, and in fact most often, of the order of the Planck length \cite{ST2}. 
In this section, we examine
whether compact dimensions have an effect on the BEC critical temperature, in which case they may be measurable. 
Interestingly, we find that there is indeed 
such an effect. We start with the expression for the 
charge density $n$ of relativistic particles in $d$ non-compact dimensions and $N$ compact dimensions with a topology of $\mathbb{R}^d\times S^N$ \cite{shiraishi}, 
%
%
\begin{eqnarray}
\label{cd}
n=\sum_{\ell=0}^\infty d_\ell\int_0^\infty \frac{\mathrm{d}^dk}{(2\pi)^d}\left[\frac{1}{e^{\beta\left(\sqrt{\hbar^2k^2c^2+m^2c^4+\hbar^2\omega_\ell^2}-\mu\right)}-1}-\frac{1}{e^{\beta\left(\sqrt{\hbar^2k^2c^2+m^2c^4+\hbar^2\omega_\ell^2}+\mu\right)}-1}\right]~,
\end{eqnarray}
where
\begin{eqnarray}
\label{cdparams}
d_\ell \equiv \frac{(2\ell+N-1)\Gamma(\ell+N-1)}{\ell!\,\Gamma(N)}\,\,\,\,\,\,\mathrm{and}\,\,\,\,\,\,\omega_\ell^2 \equiv \frac{c^2}{R^2}\,\ell(\ell+N-1)\,\,\,\,\,\,
(\mathrm{for}\,\,\,\ell\in\mathbb{N}\cup\{0\})~,
\end{eqnarray}
are the degeneracy factors and energy contributions from compact dimensions respectively, and $R$ is the radius of the compact $S^N$. 
We consider the case which is currently the only one which is experimentally measurable, 
namely a non-relativistic BEC, with $k_BT\ll mc^2$. In this case, Eq.(\ref{cd}) reduces to 
\begin{eqnarray}
\label{cdn}
n\simeq\sum_{\ell=0}^\infty d_\ell\frac{1}{(2\pi\hbar^2)^{d/2}}(k_BT_c)^{d/2}\frac{1}{c^d}\left(\sqrt{\hbar^2\omega_\ell^2+m^2c^4}\right)^{d/2}\sum_{n=1}^\infty\frac{1}{n^{d/2}}e^{-n\beta_c\left(\sqrt{\hbar^2\omega_\ell^2+m^2c^4}-mc^2\right)}~.
\end{eqnarray}
If we further consider a regime where the radius of the compact dimension is very small, and the boson mass is much less than its inverse mass scale (the Planck scale), such that the inequality $(mc/h)R\ll 1$ holds, 
we see that all terms except for $\ell=0$ are exponentially suppressed by the Boltzmann factor. The second sum in Eq.(\ref{cdn}) reduces to the polylogarithm function $Li_{d/2}(\exp{(-\beta_c\tfrac{\hbar c}{R}\sqrt{N})})$ (see Eq.(\ref{plf}) in Appendix \ref{mtsm}), where we have used the condition 
$\hbar\omega_\ell\gg mc^2$, which follows directly from the previous one for small mass regimes. 
Now, for a small argument of the polylogarithm, one can write $Li_{d/2}(\exp{(-\beta_c\tfrac{\hbar c}{R}\sqrt{N})}) \approx \exp{(-\beta_c\tfrac{\hbar c}{R}\sqrt{N})}$ \cite{PLF}. Using this, with additionally including only the $\ell=1$ term, we get the number density of bosons including the correction term due to compact dimensions from Eq.(\ref{cdn})
\begin{eqnarray}
\label{bendc}
n\simeq\left(\frac{mk_BT_c}{2\pi\hbar^2}\right)^{d/2}\left[\zeta(\tfrac{d}{2})+\frac{\hbar^{d/2}N^{({d+4})/{4}}}{R^{d/2}\,m^{d/2}\,c^{d/2}}e^{-\beta_c\tfrac{\hbar c}{R}\sqrt{N}}\right]~.
\end{eqnarray}
We can see from the above that the first term agrees with standard theory,
while the second term is the 
remnant from the extra dimensions, 
which as expected is very small due to the Boltzmann suppression factor. This term goes to zero as $R\rightarrow 0$. 
We are interested in the critical temperature $T_c$, which we extract from Eq.(\ref{bendc}), using a perturbative approach and defining 
$T_c=T_c^{(0)}+\Delta T(R)$. 
The critical temperature with corrections due to extra compact dimensions then takes the form
\begin{eqnarray}
\label{cdtc}
T_c\simeq\frac{2\pi\hbar^2}{k_B\,m\,\zeta(\tfrac{d}{2})^{2/d}}\,n^{2/d}-\frac{4\pi\hbar^{(d+4)/2}N^{({d+4})/{4}}\,e^{-\beta_c^{(0)}\tfrac{\hbar c}{R}\sqrt{N}}}{k_B\,d\,R^{d/2}\,m^{(d+2)/2}\,c^{d/2}\,\zeta(\tfrac{d}{2})^{{(2+d)}/{d}}}\,n^{2/d}~,
\end{eqnarray}
where $\beta_c^{(0)}=1/(k_BT_c^{(0)})$. From Eq.(\ref{cdtc}), we see that the first term is identical to Eq.(\ref{edtc}) and the magnitude of the correction term increases with increasing number density and decreasing boson mass. We can also see the non-trivial dependence of the correction term on the compact dimension. We discuss this in terms of the relative magnitude of the correction, expressible from 
Eq.(\ref{cdtc}) as
\begin{eqnarray}
\label{cdrc}
\left|\frac{\Delta T(R)}{T_c^{(0)}}\right|=\frac{2\,\hbar^{d/2}N^{({d+4})/{4}}\,e^{-\beta_c^{(0)}\tfrac{\hbar c}{R}\sqrt{N}}}{d\,R^{d/2}\,m^{d/2}\,c^{d/2}\,\zeta(\tfrac{d}{2})}
\equiv 10^{-r} < 10^{-q}~.
\end{eqnarray}
%
In the above, $r$ and $q$ take positive values and
$10^{-q}$ denotes the precision at which the BEC critical temperature can be measured, and the inequality stems from the fact that the above $\Delta T(R)$ has not been observed in the laboratory so far. This subsequently puts bounds on the extra dimensions, as we shall see below. The important point to note here is that the RHS of Eq.(\ref{cdrc}) contains the compact dimension radius $R$ in the denominator {\it as well as} in the numerator, via the exponential Boltzmann factor.  
Therefore, interestingly, as one spans the range of $R$ from very small to larger values, the correction term first increases and then starts to decrease. This behaviour is shown in Fig.(\ref{cdfig}). 
The blue line therein depicts the relative correction
given in Eq.(\ref{cdrc}), suitably normalized for ease of comparison with the horizontal orange line, signifying a hypothetical precision, expected to be attainable in the future (a line corresponding to current accuracies would lie well above the blue curve). 
%
%
Note that the curves intersect at two points, corresponding to $R_1$ and $R_2$ on the horizontal axis.
Therefore, if no trace of the compact dimension is found in experiments, in terms of the above corrections, it would mean that the relative precision is either on the left of $R_1$ or right of $R_2$. In other words, one obtains an upper {\it as well as} a lower bound on the size of the compact radius $R$.
%
%
More precisely, 
the peak is located at $R_\mathrm{max}=2\beta_c^{(0)}\hbar c\sqrt{N}/d$ and the two bounds always satisfy $R_1<R_\mathrm{max}$ and $R_2>R_\mathrm{max}$ for lower and upper bound respectively. 
While upper bounds on the size of compact dimensions have been imposed from a number of theoretical and experimental standpoints, 
We are not aware of any other experiment or observation which puts both an upper {\it and} a lower bounds on $R$. 
We can also see that Eq.(\ref{cdrc}) is implicitly dependent on the number density $n$ through the 
inverse critical temperature 
$\beta_c^{(0)}$, since the latter depends on $n$. Therefore, if we increase $n$, the exponential factor increases and the relative correction increases.
%
%
Therefore, the magnitude of the correction, signified by the quantity $r$, which one hopes to minimise, is given in terms of $R$, the number of compact dimensions $N$, the boson mass $m$ and implicitly on the boson density $n$ as 
%
%
\begin{eqnarray}
r=r_{d,N}(R,m)=\log{\left(\frac{d\,R^{d/2}\,m^{d/2}\,c^{d/2}\,\zeta(\tfrac{d}{2})}{2\,\hbar^{d/2}N^{({d+4})/{4}}}\,e^{\beta_c^{(0)}\tfrac{\hbar c}{R}\sqrt{N}}\right)}~.
\end{eqnarray}
The above gives us an estimate of what the precision of the temperature measurements must be, to detect such a deviation from the standard theory and obtain signatures of extra dimensions. 
The second observable, the fraction of bosons in the ground state, including corrections due to extra compact dimensions, takes the form
\begin{eqnarray}
\label{cdfr}
f_0=\frac{n_0}{n}=1-\left(\frac{T}{T_c}\right)^{d/2}\left[1+\frac{\hbar^{d/2}N^{(d+4)/4}}{R^{d/2}\,m^{d/2}\,c^{d/2}\,\zeta(\tfrac{d}{2})}\left(e^{-\beta_c\tfrac{\hbar c}{R}\sqrt{N}\left(\tfrac{T_c}{T}\right)}-e^{-\beta_c\tfrac{\hbar c}{R}\sqrt{N}}\right)\right]~,
\end{eqnarray}
from where we can see, that without the correction term ($R\rightarrow0$) the fraction is the same as in standard theory in Eq.(\ref{fderiv}). The correction term in square brackets in Eq.(\ref{cdfr}) vanishes for $T=T_c$, as expected. 


In this analysis, we considered a spherical topology of the compact dimensions $\mathbb{R}^d\times S^N$, where there is only one radius, no matter how many compact dimensions $N$ we consider. If the topology of the compact dimensions is toroidal instead, say $\mathbb{R}^d\times T(N)$, then each of the $N$ compact dimensions could have distinct radii. For such a spatial topology, the above calculations would be similar, with the difference that the energy contribution in Eq.(\ref{cdparams}) would be a function of all $N$ radii, instead of just $R$. However, for $N=1$, we do not expect any difference between the two topologies.

\section{Bose-Einstein condensation with QG corrections}
\label{QGBEC}

QG effects in standard QM systems are implied by the GUP, defined by the commutator in Eq.(\ref{gup}). The phase space integrals in statistical mechanics are normalized by a phase space volume of a particle in a box, so this is where the QG corrections appear in the analysis. Such a modification also modifies the density of states, which is used to calculate the QG corrected number density of bosons and by extension, the critical temperature and fraction of bosons in the ground state. To apply the GUP in Eq.(\ref{gup}) to a quantum particle in a box, we modify the dispersion relation between energy and momentum of a particle, which is the Hamilton operator in QM
 \begin{equation}
 \label{hamilton}
     H
     =\frac{p^2}{2m}+V(\mathbf{x})~,
 \end{equation}
 where $\mathbf{p}$ is the physical momentum of the particle, $p=|\mathbf{p}|$, $m$ its mass and $V(\mathbf{x})$ the potential of a particle in a box ($V(\mathbf{x})=0$ inside the box and $V(\mathbf{x})=\infty$ outside the box).
 We notice that we cannot use the standard operator for momentum $p_i\neq-i\hbar\partial_{x_i}$, because the commutation relation is modified as in Eq.(\ref{gup}). 
 However, we can define a set of canonical operators $x_{0i}$ and $p_{0i}$, which satisfy a standard commutation relation $[x_{0i},p_{0j}]=i\hbar\delta_{ij}$.
 Therefore, we can write
  $p_{0i}=-i\hbar\partial_{x_{0i}}$. In terms of 
  $x_{0i}$ and $p_{0i}$, we get 
 \begin{equation}
 \label{cops}
     x_i=x_{0i}~,\,\,\,\,\,\,\,\,p_i=p_{0i}(1-\alpha p_0+2\beta p_0^2)~,
 \end{equation}
 where $p_0=\sqrt{p_{0k}\,p_{0k}}$. 
We will use the above to compute QG corrections to the non-relativistic and relativistic Hamiltonians, and examine its consequences for a BEC in the following subsections. Note that the physical momentum is still $p_i$. 

\subsection{Non-relativistic}

In this case, we use the non-relativistic kinetic term and choose the potential inside a three dimensional box with edges $L_x$,
$L_y$ and $L_z$ to be $V(\mathbf{x})=0$, and the potential outside this box $V(\mathbf{x})=\infty$.
As usual, we choose 
the boundary conditions $\psi(0,y,z)=\psi(x,0,z)=\psi(x,y,0)=\psi(L_x,y,z)=\psi(x,L_y,z)=\psi(x,y,L_z)=0$. To compute the QG corrected energy spectrum of a non-relativistic particle in a three dimensional box, 
we first write the QG corrected Hamiltonian by replacing $p$ in terms of $p_0$ as given in Eq.(\ref{cops})
\begin{equation}
\label{hamilton1}
    H=\frac{p^2}{2m}
    =\frac{p_0^2}{2m}-\frac{\alpha}{m}p_0^3+\frac{5\beta}{2m}p_0^4 \equiv H_0+H_1+H_2~,
\end{equation}
where $H_0=\frac{p_0^2}{2m}$, $H_1=-\frac{\alpha}{m}p_0^3$ and $H_2=\frac{5\beta}{2m}p_0^4$. We will compute corrections to the energy spectrum due to $H_1$ and $H_2$ to linear order in $\beta$ and quadratic order in $\alpha$ (note that these are of a similar order or magnitude). 
As we know, the eigenfunctions of an unperturbed Hamiltonian $H_0$, for a particle in a three dimensional box are given as \cite{JLB}
\begin{equation}
\label{efpib}
    \psi_\mathbf{n}(\mathbf{x}_0)=\psi_{n_x,n_y,n_z}(x_0,y_0,z_0)=\sqrt{\frac{8}{V}}\,\sin{\left(\frac{\pi n_x}{L_x}x_0\right)}\sin{\left(\frac{\pi n_y}{L_y}y_0\right)}\sin{\left(\frac{\pi n_z}{L_z}z_0\right)}~,
\end{equation}
where $V=L_xL_yL_z$ is the volume of the box and $n_x,n_y, n_z\in\mathbb{N}$ are quantum numbers. In a Hilbert space $\mathcal{H}=\{\psi_\mathbf{n}; \mathbf{n}\in\mathbb{N}^3\}$ we can write a general wavefunction as $\Psi(\mathbf{x}_0)=\sum_\mathbf{n}c_\mathbf{n}\psi_\mathbf{n}(\mathbf{x}_0)$, where $c_\mathbf{n}\in\mathbb{C}$. The energy spectrum of a three dimensional particle in a box, considering an unperturbed Hamiltonian $H_0$ is
\begin{equation}
\label{pibe}
    \varepsilon_\mathbf{n}^{(0)}=\varepsilon_{n_x,n_y,n_z}^{(0)}=\langle\psi_\mathbf{n}(\mathbf{x}_0)|H_0|\psi_\mathbf{n}(\mathbf{x}_0)\rangle=\frac{\hbar^2\pi^2}{2mL^2}(n_x^2+n_y^2+n_z^2)~,
\end{equation}
where we assumed $L=L_x=L_y=L_z$ without loss of generality. To get the QG correction to the energy spectrum in Eq.(\ref{pibe}), we use the time independent, first order perturbation theory to compute the linear (see Appendix \ref{lincalc}) and quadratic terms of the perturbation $H_1$ and $H_2$ respectively as
\begin{eqnarray}
    \Delta\varepsilon_\mathbf{n}^{(1)Lin}\!\!\!\!&=&\!\!\!\!\langle\psi_\mathbf{n}(\mathbf{x}_0)|H_1|\psi_\mathbf{n}(\mathbf{x}_0)\rangle=-\frac{\alpha\hbar^3\pi^3}{mL^3}(n_x^2+n_y^2+n_z^2)^{3/2}
    \label{pibe1}\\
    \Delta\varepsilon_\mathbf{n}^{(1)Quad}\!\!\!\!&=&\!\!\!\!\langle\psi_\mathbf{n}(\mathbf{x}_0)|H_2|\psi_\mathbf{n}(\mathbf{x}_0)\rangle=\frac{5\beta\hbar^4\pi^4}{2mL^4}(n_x^4+n_y^4+n_z^4+2n_x^2n_y^2+2n_x^2n_z^2+2n_y^2n_z^2)~,
    \label{pibe2}
\end{eqnarray}
so that the energy spectrum of a particle in a three dimensional box, up to quadratic order of the QG parameters, is just the sum of Eqs.(\ref{pibe}-\ref{pibe2})
\begin{eqnarray}
\label{qgpib}
    \varepsilon_\mathbf{n}\!\!\!\!&=&\!\!\!\!\frac{\hbar^2\pi^2}{2mL^2}(n_x^2+n_y^2+n_z^2)-\frac{\alpha\hbar^3\pi^3}{mL^3}(n_x^2+n_y^2+n_z^2)^{3/2}  \nonumber \\ 
    &+&\!\!\!\!\frac{5\beta\hbar^4\pi^4}{2mL^4}(n_x^4+n_y^4+n_z^4+2n_x^2n_y^2+2n_x^2n_z^2+2n_y^2n_z^2) \nonumber \\
    &=&\!\!\!\!\frac{\hbar^2}{2m}k_\mathbf{n}^2-\frac{\alpha\hbar^3}{m}k_\mathbf{n}^3+\frac{5\beta\hbar^4}{2m}k_\mathbf{n}^4~,
\end{eqnarray}
where $k_\mathbf{n}^2=\frac{\pi^2}{L^2}\left(n_x^2+n_y^2+n_z^2\right)$ in the third line. From the above we can see that the QG corrections to the energy spectrum of a particle in a three dimensional box are also dependent on quantum numbers $n_x$, $n_y$ and $n_z$, but with different powers. An exact procedure to obtain the QG corrected energy spectrum of a particle in a one dimensional box, without using perturbation theory is described in \cite{pbpb}.

 Considering the QG corrected energy spectrum for a particle in a three dimensional box, given by Eq.(\ref{qgpib}), we calculate the QG corrected density of states in the continuum limit $\varepsilon_\mathbf{n}\longrightarrow\varepsilon$ (see Appendix \ref{dp}) as
\begin{eqnarray}
\label{qgdos}
g(\varepsilon)=\frac{V(2m)^{3\slash2}\varepsilon^{1\slash2}}{4\pi^2\hbar^3}(1+16\alpha\sqrt{m}\,\varepsilon^{1/2}-25\beta m\varepsilon)~,
\end{eqnarray}
from which we see that it reduces to the usual density of states, given by Eq.(\ref{dos}) in Appendix \ref{mtsm}, when $\alpha,\beta\rightarrow0$. 
A more general procedure, described in Appendix \ref{RKP} can also be used to obtain the QG corrected density of states. However, its use is currently limited to the non-relativistic case with quadratic corrections only, due to the complexity of calculations.  

\begin{figure}[H]
     \centering
     \includegraphics[scale=0.85]{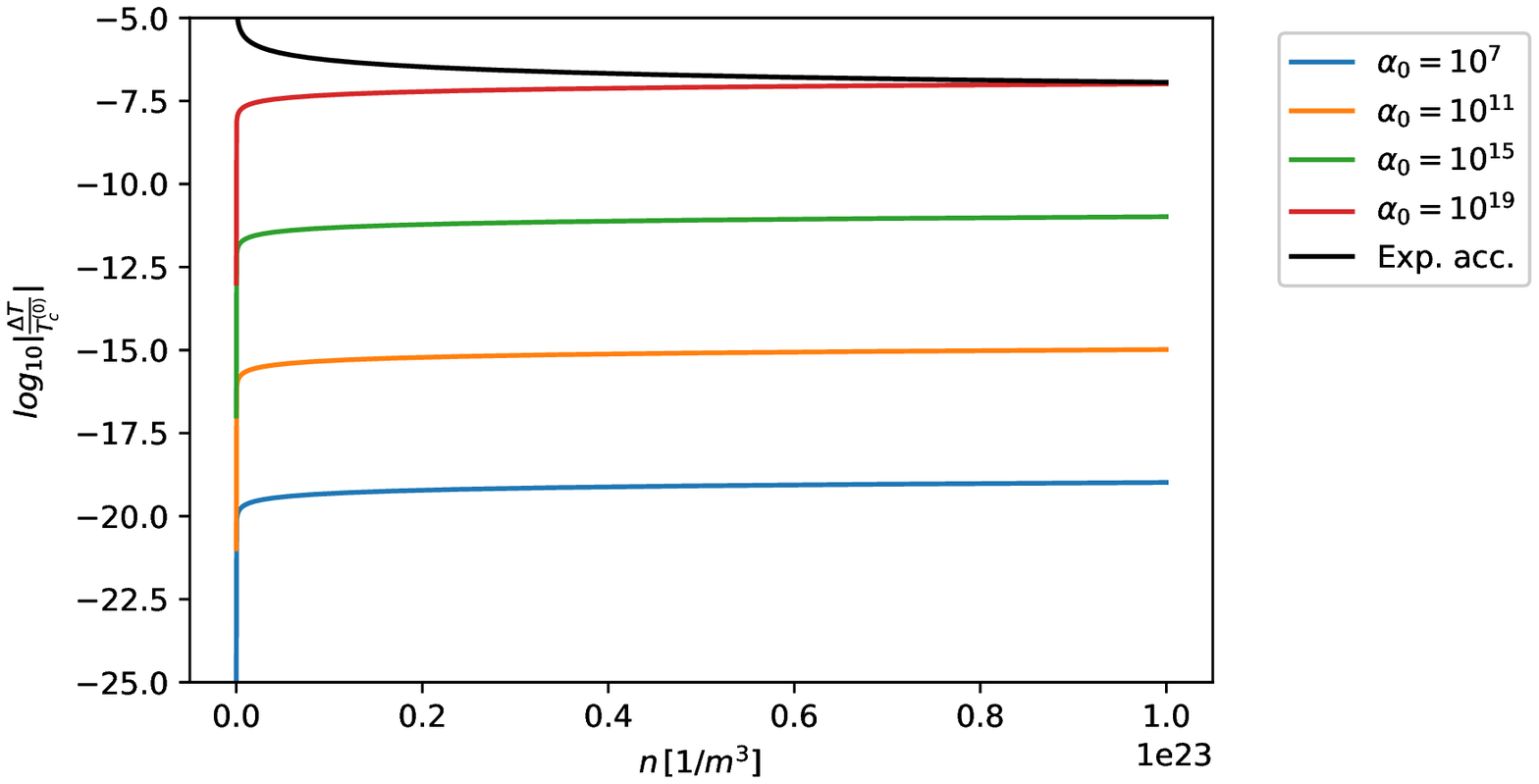}
     \caption{Relative temperature correction as a function of the number density $n$, for a helium gas, for different values of parameter $\alpha_0$, where $\beta_0=\alpha_0^2$ and the black line represents the experimental accuracy.}
     \label{qgnrfig}
\vspace{0.6cm}
     \centering
     \includegraphics[scale=0.85]{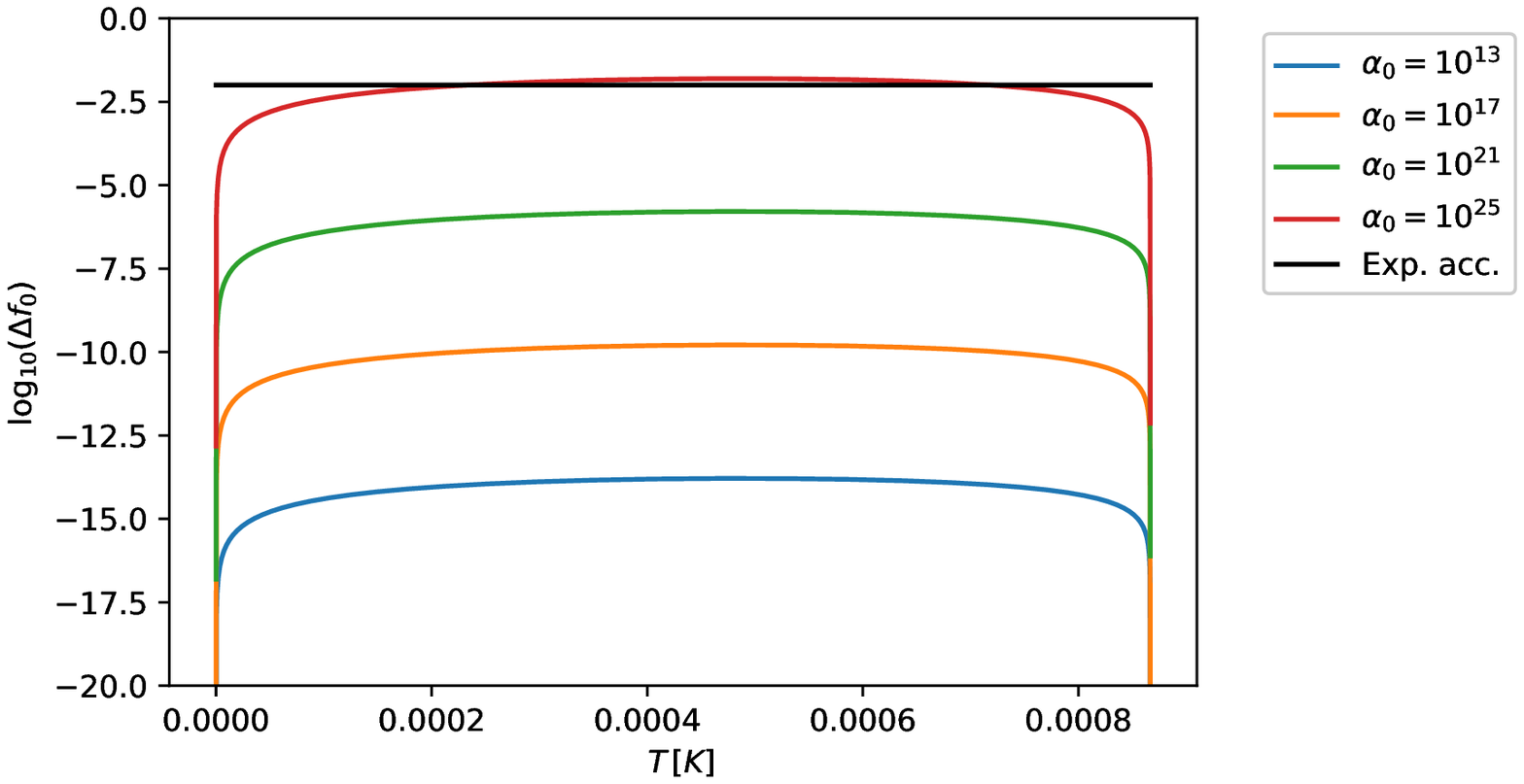}
     \caption{Fraction correction as a function of the condensate temperature $T$, for a helium gas, for different values of parameter $\alpha_0$, where $\beta_0=\alpha_0^2$ and the black line represents the experimental accuracy.}
     \label{fracn}
 \end{figure}

The number of particles in the system is calculated using Eq.(\ref{ensavg}) of Appendix \ref{mtsm} and the QG corrected density of states in Eq.(\ref{qgdos}). We evaluate the integral at $T_c$ ($\mu\rightarrow0$) and divide it by $V$ to get the QG corrected boson number density
\begin{eqnarray}
\label{qgnd}
n\!\!\!\!&=&\!\!\!\!\frac{N_{\,\!_{BE}}}{V}=\frac{\sqrt{2}m^{3/2}}{2\pi^2\hbar^3}\left[\int_0^\infty\frac{\varepsilon^{1/2}}{e^{\beta_c\varepsilon}-1}\mathrm{d}\varepsilon+16\alpha\sqrt{m}\int_0^\infty\frac{\varepsilon}{e^{\beta_c\varepsilon}-1}\mathrm{d}\varepsilon-25\beta m\int_0^\infty\frac{\varepsilon^{3/2}}{e^{\beta_c\varepsilon}-1}\mathrm{d}\varepsilon\right] \nonumber \\
&=&\!\!\!\!\frac{\sqrt{2}m^{3/2}}{4\pi^{3/2}\hbar^3}\left[(k_BT_c)^{3/2}\zeta(\tfrac{3}{2})+\frac{16\pi^{3/2}}{3}\alpha\sqrt{m}(k_BT_c)^2-\frac{75}{2}\beta m(k_BT_c)^{5/2}\zeta(\tfrac{5}{2})\right]~,
\end{eqnarray}
where we again see that it reduces to the usual number density, given by Eq.(\ref{nbe}) (as $\mu\rightarrow0$) in Appendix \ref{mtsm}, when $\alpha,\beta\rightarrow0$. Note that we cannot extract a closed form expression of 
$T_c$ from Eq.(\ref{qgnd}), and therefore we
use a perturbative approach. 
We define $T_c=T_c^{(0)}+\Delta T(\alpha)+\Delta T(\beta)$,
to express the QG corrected $T_c$, where $\Delta T(\alpha)\propto \alpha$ and $\Delta T(\beta)\propto \beta$. It is easy to see that the uncorrected critical temperature 
$T_c^{(0)}$ is equal to that in Eq.(\ref{edtc}) for $d=3$. 
The QG corrected critical temperature $T_c$ is then
\begin{eqnarray}
\label{qgtcnr}
T_c=\frac{2\pi\hbar^2}{k_Bm\zeta(\tfrac{3}{2})^{2/3}}n^{2/3}-\alpha\frac{32\sqrt{8}\pi^3\hbar^3}{9k_Bm\zeta(\tfrac{3}{2})^2}n+\beta\frac{100\pi^2\hbar^4\zeta(\tfrac{5}{2})}{k_Bm\zeta(\tfrac{3}{2})^{7/3}}n^{4/3}~,
\end{eqnarray}
where we can see that the QG corrections increase with increasing number density $n$ and decreasing boson mass $m$. We also see that higher order QG corrections have a stronger dependence on $n$.
This is a direct consequence of the presence of 
higher order terms $T_c$ in Eq.(\ref{qgnd}). 
The magnitude of the relative correction is then
\begin{eqnarray}
\label{rqgnr}
\left|\frac{\Delta T}{T_c^{(0)}}\right|=\alpha_0\frac{16\sqrt{8}\pi^2\hbar}{9M_Pc\zeta(\tfrac{3}{2})^{4/3}}n^{1/3}-\beta_0\frac{50\pi\hbar^2\zeta(\tfrac{5}{2})}{(M_Pc)^2\zeta(\tfrac{3}{2})^{5/3}}n^{2/3}~,
\end{eqnarray}
which increases with increasing $n$, 
but does not depend on the boson mass. 
This is presented in Fig(\ref{qgnrfig}), where the black line represents the current experimental accuracy which will evidently
continue to improve with time. Eq.(\ref{rqgnr}) differs from 
a similar result in \cite{BGL,FB}, where the relative correction decreases with increasing $n$ as $|\Delta T/T_c^{(0)}|\propto\alpha_0/n^{1/3}$. 
Note that as the particle number increases in a given volume, 
the total energy gets closer to the Planck energy scale, thus magnifying the QG effects \cite{GAC2}. This shows that our result is perfectly reasonable.


The second important observable is the fraction of bosons in the ground state. Using the same procedure as for Eq.(\ref{fderiv}), we calculate this
fraction using Eq.(\ref{qgnd}) as
\begin{eqnarray}
\label{frnr}
f_0=\frac{n_0}{n}=1-\left(\frac{T}{T_c}\right)^{3/2}+\alpha\frac{16\pi^{3/2}}{3\zeta(\tfrac{3}{2})}\sqrt{mk_B}\left[\frac{T^{3/2}}{T_c}-\frac{T^2}{T_c^{3/2}}\right]-\beta\frac{75}{2}\frac{\zeta(\tfrac{5}{2})}{\zeta(\tfrac{3}{2})}mk_B\left[\frac{T^{3/2}}{T_c^{1/2}}-\frac{T^{5/2}}{T_c^{3/2}}\right]~,
\end{eqnarray}
where we see that a standard result in Eq.(\ref{fderiv}) is recovered for $\alpha,\beta\rightarrow 0$. 
Furthermore, we see that at $T=T_c$, the QG corrections vanish and $f_0=0$, as expected, even when $\alpha,\beta\neq0$. 
This means that a deviation in fraction of bosons in the ground state due to QG effects, should be observed at temperatures $T<T_c$. 
 The corrections terms from Eq.(\ref{frnr}) are presented in Fig.(\ref{fracn}) for a helium gas. We see that the  correction has a maximum between absolute zero and the critical temperature $T_c$ at
 \begin{eqnarray}
 T_m=\frac{9}{16}T_c-\frac{\beta}{\alpha}\frac{2025}{256}\frac{\zeta(\tfrac{5}{2})\sqrt{mk_B}}{\pi^{3/2}}T_c^{3/2}~.
 \end{eqnarray}
 This suggest that experiments able to measure the fraction of bosons in the ground state with high accuracy would most likely observe QG corrections near this temperature $T_m$.

\subsection{Relativistic}

For this case we follow the procedure outlined in \cite{ADV} 
to obtain the QG corrected energy spectrum of a relativistic boson in a three dimensional box.
This is then used to obtain the relativistic density of states, the QG corrected critical temperature $T_c$, and fraction of bosons in the ground state $f_0$ for a relativistic BEC. We consider two distinct cases. In the first case, we consider neutral bosons, and in the second case we consider charged bosons and antibosons.

\begin{figure}[H]
     \centering
     \hbox{\hspace{-0.4cm}\includegraphics[scale=0.85]{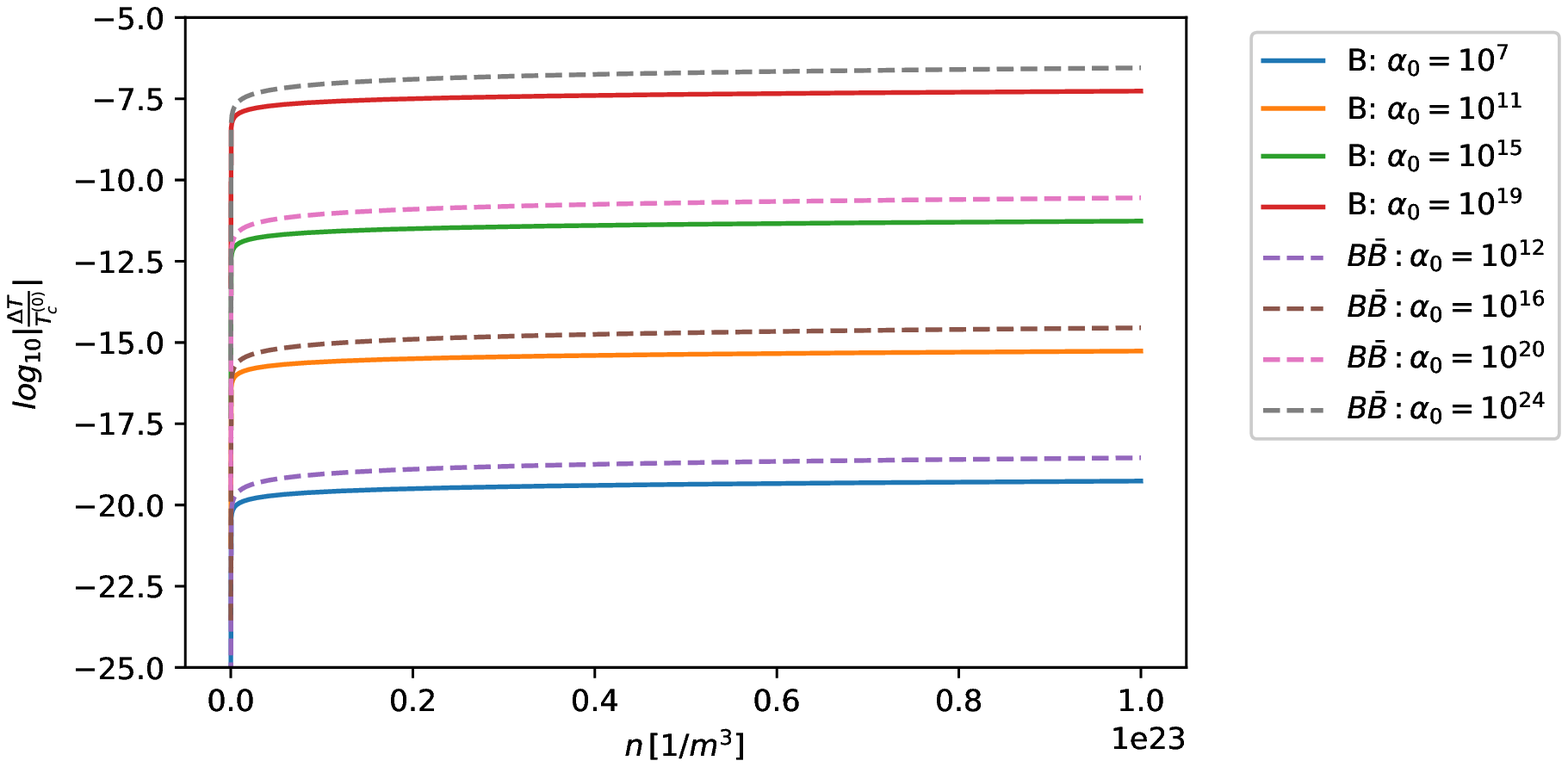}}
     \caption{Relative correction as a function of the number density $n$, for a helium gas, for different values of parameter $\alpha_0$, where $\beta_0=\alpha_0^2$. The solid lines represent the neutral case and the dashed lines represent the charged case.}
     \label{qgrfig}
  \vspace{0.6cm}
  \centering
     \includegraphics[scale=0.85]{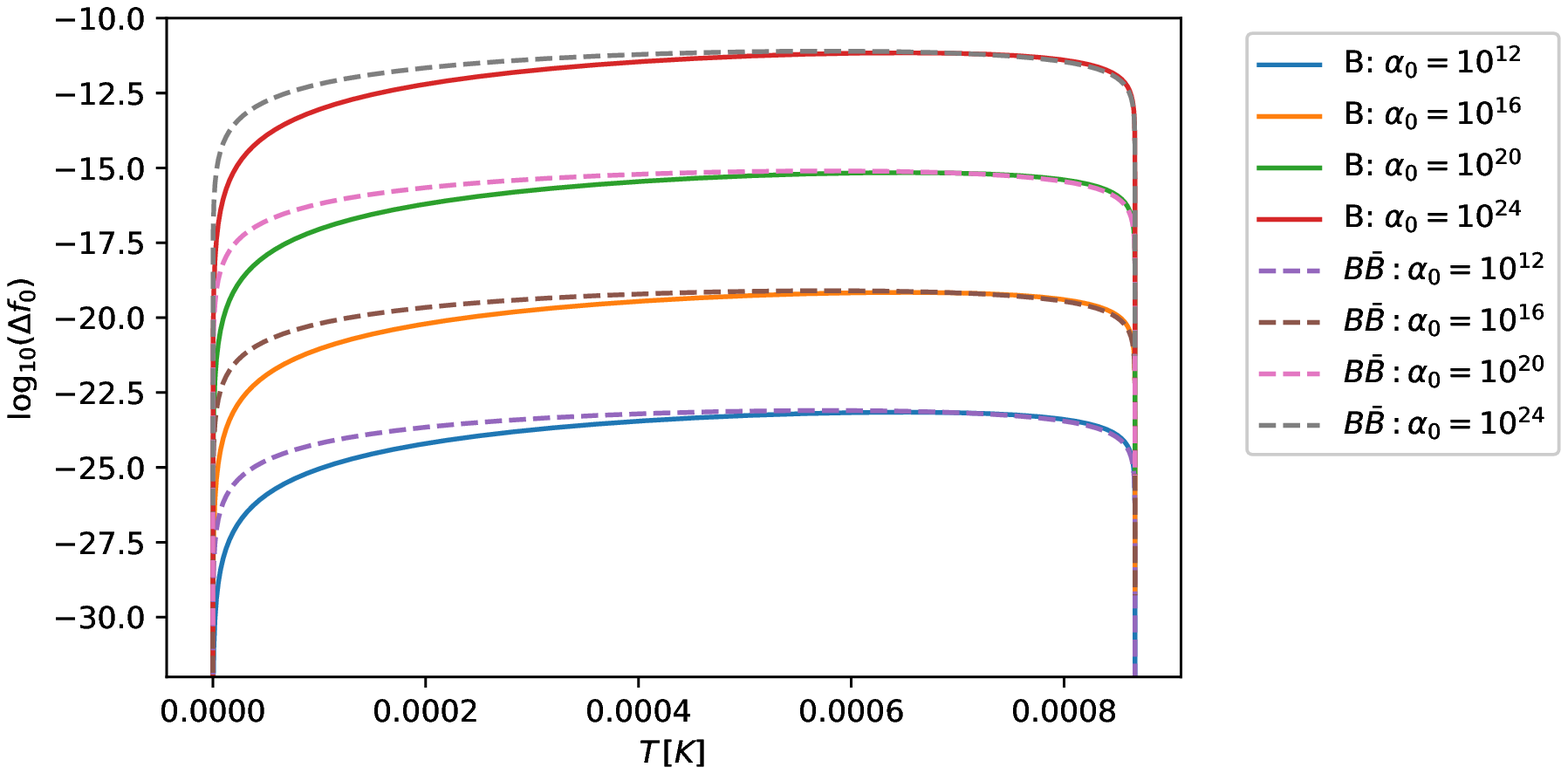}
     \caption{Fraction correction as a function of the condensate temperature $T$, for a helium gas, for different values of parameter $\alpha_0$, where $\beta_0=\alpha_0^2$. The solid lines represent the neutral case and the dashed lines represent the charged case.}
     \label{fracr}
 \end{figure}

Relativistic bosons (spin zero) are described by the Klein-Gordon equation, which as shown in \cite{ADV, FVF}
gives rise to the following effective Hamiltonian in the Feshbach-Villars formalism
%
%
\begin{eqnarray}
\label{relhamil}
H=(\tau_3+i\tau_2)\frac{\mathbf{p}^2}{2m}+\tau_3\, mc^2~,
\label{ham1}
\end{eqnarray}
where $\tau_i~(i=1,2,3)$ are the Pauli matrices.
The corresponding wavefuction satisfies the  
equation $i\hbar\,\partial_t\Psi=H\Psi$. The eigenfunctions $\Psi$ of the effective Hamiltonian in Eq.(\ref{relhamil}), are given by
\begin{eqnarray}
\Psi=\left(\begin{array}{c}
     \varphi  \\
      \chi
\end{array}\right)~.
\end{eqnarray}
It is straightforward to show that $\psi=\varphi+\chi$ satisfies the Klein-Gordon equation
\begin{eqnarray}
\frac{1}{c^2}\frac{\partial^2\psi}{\partial t^2}-\nabla^2\psi+\frac{m^2c^2}{\hbar^2}=0~.
\end{eqnarray}
It turns out that the time dependent solutions for $\varphi$ and $\chi$ of the effective Hamiltonian 
in Eq.(\ref{ham1}) are
\begin{eqnarray}
\varphi_\mathbf{n}^\pm(\mathbf{x})\!\!\!\!&=&\!\!\!\!\sqrt{\frac{8}{V}}\varphi_0^\pm(\mathbf{p})e^{\mp\frac{E}{\hbar}t}\sin{\left(\frac{\pi n_x}{L_x}x\right)}\sin{\left(\frac{\pi n_y}{L_y}y\right)}\sin{\left(\frac{\pi n_z}{L_z}z\right)}\,\,\,\,\,\,\, \nonumber \\
\chi_\mathbf{n}^\pm(\mathbf{x})\!\!\!\!&=&\!\!\!\!\sqrt{\frac{8}{V}}\chi_0^\pm(\mathbf{p})e^{\mp\frac{E}{\hbar}t}\sin{\left(\frac{\pi n_x}{L_x}x\right)}\sin{\left(\frac{\pi n_y}{L_y}y\right)}\sin{\left(\frac{\pi n_z}{L_z}z\right)}~,
\label{fihi}
\end{eqnarray}
where $V=L_xL_yL_z$ is the volume of the box, $n_x,n_y, n_z\in\mathbb{N}$ are quantum numbers and $(\varphi_0^\pm)^2-(\chi_0^\pm)^2=\pm1$ where $\pm$ denotes particle and anti-particle solutions. We notice that the solutions in Eq.(\ref{fihi}) are similar to those in the non-relativistic case in Eq.(\ref{efpib}). They differ only by the relativistic, momentum-dependent functions $\varphi_0^\pm(\mathbf{p})$ and $\chi_0^\pm(\mathbf{p})$. To compute the QG corrected energy spectrum of a relativistic particle in a box, we modify the effective Hamiltonian, using the transformation in Eq.(\ref{cops}), as 
\begin{eqnarray}
\label{QGhamil}
H_{QG}=(\tau_3+i\tau_2)\frac{p_0^2}{2m}+\tau_3mc^2 -(\tau_3+i\tau_2)\frac{\alpha}{m}p_0^3+(\tau_3+i\tau_2)\frac{5\beta}{2m}p_0^4=H_0+H_1+H_2~,
\end{eqnarray}
where $H_0=(\tau_3+i\tau_2)\frac{p_0^2}{2m}+\tau_3mc^2$, $H_1=-(\tau_3+i\tau_2)\frac{\alpha}{m}p_0^3$ and $H_2=(\tau_3+i\tau_2)\frac{5\beta}{2m}p_0^4$. The energy spectrum of a three dimensional relativistic particle in a box, considering an unperturbed effective Hamiltonian $H_0$ is
\begin{eqnarray}
\label{piber}
\varepsilon_{\mathbf{n}}^{(0)}=\pm\sqrt{\frac{c^2\hbar^2\pi^2}{L^2}\left(n_x^2+n_y^2+n_z^2\right)+m^2c^4}~,
\end{eqnarray}
where we assumed $L=L_x=L_y=L_z$, without loss of generality and $\pm$ signifies the particle and anti-particle solutions.
We obtained the energy spectrum in Eq.(\ref{piber}), by computing the eigenvalues of the $H_0$ operator.
%
To get the QG correction to the energy spectrum in Eq.(\ref{piber}), we consider the complete QG corrected, effective Hamiltonian $H_{QG}$ from Eq.(\ref{QGhamil}) and use the result from Appendix \ref{lincalc}. The QG corrected energy spectrum is then
\begin{eqnarray}
\label{qgpibr}
\varepsilon_\mathbf{n}=\pm\sqrt{\hbar^2c^2k_\mathbf{n}^2-2\alpha\hbar^3c^2k_\mathbf{n}^3+5\beta\hbar^4c^2k_\mathbf{n}^4+m^2c^4}~,
\end{eqnarray}
which is obtained by calculating the eigenvalues of the effective Hamiltonian $H_{QG}$.
%
In the above, we again used $k_\mathbf{n}^2=\frac{\pi^2}{L^2}\left(n_x^2+n_y^2+n_z^2\right)$ and the sign plays no role in further considerations, since we use the square of Eq.(\ref{qgpibr}). Considering the QG corrected energy spectrum for a particle in a three dimensional box, given by the relativistic relation in 
Eq.(\ref{qgpibr}), we calculate the QG corrected density of states in the continuum limit $k_\mathbf{n}\longrightarrow k$ and  $\varepsilon_\mathbf{n}\longrightarrow\varepsilon$ (see Appendix \ref{dp}) as
\begin{eqnarray}
\label{qgdosr}
g(\varepsilon)=\frac{V\varepsilon\sqrt{\varepsilon^2-m^2c^4}}{2\pi^2\hbar^3c^3}\left(1+4\alpha\frac{1}{c}\sqrt{\varepsilon^2-m^2c^4}-\frac{25}{2}\beta\frac{1}{c^2}\left(\varepsilon^2-m^2c^4\right)\right)~.
\end{eqnarray}
We see that it reduces to the usual relativistic density of states, given by Eq.(\ref{dosr}) in Appendix \ref{mtsm}, when $\alpha,\beta\rightarrow0$. It may be noted that the integrals which take the form of Eq.(\ref{ensavg}) in Appendix \ref{mtsm} are non-analytical when
using the relativistic density of states in Eq.(\ref{qgdosr}). 
They can be expressed in a closed form only in the ultra-relativistic (UR) limit, where $\varepsilon\gg mc^2$. The number of particles in the system is calculated using Eq.(\ref{ensavg}) in Appendix \ref{mtsm} and the QG corrected density of states in Eq.(\ref{qgdosr}). We evaluate the integral at $T_c$ ($\mu\rightarrow0$) in the UR limit and divide it by $V$ to get the QG corrected number density for the neutral boson case
\begin{eqnarray}
n\!\!\!\!&=&\!\!\!\!\frac{N_{\,\!_{BE}}^{\,\!^{UR-B}}}{V}=\frac{1}{2\pi^2\hbar^3c^3}\left[\int_0^\infty\frac{\varepsilon^2}{e^{\beta_c\varepsilon}-1}\mathrm{d}\varepsilon+4\frac{\alpha}{c}\int_0^\infty\frac{\varepsilon^3}{e^{\beta_c\varepsilon}-1}\mathrm{d}\varepsilon-\frac{25}{2}\frac{\beta}{c^2}\int_0^\infty\frac{\varepsilon^4}{e^{\beta_c\varepsilon}-1}\mathrm{d}\varepsilon\right] \nonumber \\
&=&\!\!\!\!\frac{1}{\pi^2\hbar^3c^3}\left[(k_BT_c)^3\zeta(3)+\frac{2\pi^4}{15}\frac{\alpha}{c}(k_BT_c)^4-150\frac{\beta}{c^2}(k_BT_c)^5\zeta(5)\right]
\end{eqnarray}
and for the charged boson case
\begin{eqnarray}
n\!\!\!\!&=&\!\!\!\!\frac{N_{\,\!_{BE}}^{\,\!^{UR-B\bar{B}}}}{V}=\frac{m}{2\pi^2\hbar^3ck_BT_c}\left[\int_0^\infty\frac{\varepsilon^2}{\cosh{(\beta_c\varepsilon)}-1}\mathrm{d}\varepsilon+4\frac{\alpha}{c}\int_0^\infty\frac{\varepsilon^3}{\cosh{(\beta_c\varepsilon)}-1}\mathrm{d}\varepsilon\right. \nonumber \\
&-&\!\!\!\!\left.\frac{25}{2}\frac{\beta}{c^2}\int_0^\infty\frac{\varepsilon^4}{\cosh{(\beta_c\varepsilon)}-1}\mathrm{d}\varepsilon\right] \nonumber \\
&=&\!\!\!\!\frac{m}{3\hbar^3c}\left[(k_BT_c)^2+\frac{72}{\pi^2}\frac{\alpha}{c}(k_BT_c)^3\zeta(3)-10\pi^2\frac{\beta}{c^2}(k_BT_c)^4\right]~,
\end{eqnarray}
where we see, that for both cases the results return the number densities as in the standard theory \cite{GLB}, when $\alpha,\beta\rightarrow0$. Again using the  perturbative approach as for the non-relativistic case, we find the critical temperatures for the neutral boson case to be
\begin{eqnarray}
\label{qgtcbr}
T_c^B=\frac{\pi^{2/3}\hbar c}{k_B\zeta(3)^{1/3}}n^{1/3}-\alpha\frac{2}{45}\frac{\pi^{16/3}\hbar^2c}{k_B\zeta{(3)}^{5/3}}n^{2/3}+\beta\,50\frac{\pi^2\hbar^3 c}{k_B}\frac{\zeta(5)}{\zeta(3)^2}n
\end{eqnarray}
and for the charged boson case, it is given by
\begin{eqnarray}
\label{qgtcbbr}
T_c^{B\bar{B}}=\frac{1}{k_B}\left(\frac{3\hbar^3 c}{m}\right)^{1/2}n^{1/2}-\alpha108\frac{\hbar^3\zeta(3)}{\pi^2k_Bm}n+\beta\,15\frac{\pi^2}{k_B}\left(\frac{3\hbar^9}{m^3c}\right)^{1/2}n^{3/2}~,
\end{eqnarray}
where we can see that the QG corrections increase with increasing number density $n$ for both cases. In the charged boson case the QG corrections increase with decreasing boson mass $m$, while the neutral boson case is independent of boson mass. We also notice that higher order QG corrections have a stronger dependence on $n$, as also seen in the non-relativistic case. For $\alpha,\beta\rightarrow0$ in Eqs.(\ref{qgtcbr},\ref{qgtcbbr}), we recover the standard results from Eqs.(\ref{rbtc},\ref{rbbtc}). The magnitude of the relative corrections of the critical temperature for the neutral boson case is
\begin{eqnarray}
\label{rqgbr}
\left|\frac{\Delta T^B}{T_c^{(0)}}\right|=\alpha_0\frac{2\pi^{14/3}\hbar}{45M_Pc\zeta(3)^{4/3}}n^{1/3}-\beta_0\frac{50\pi^{1/2}\hbar^2\zeta(5)}{(M_Pc)^2\zeta(3)^{5/3}}n^{2/3}~,
\end{eqnarray}
while for the charged boson case, it is given by
\begin{eqnarray}
\label{rqgbbr}
\left|\frac{\Delta T^{B\bar{B}}}{T_c^{(0)}}\right|=\alpha_0\frac{108\hbar^{3/2}\zeta(3)}{\sqrt{3}\pi^2M_Pc\sqrt{mc}}n^{1/2}-\beta_0\frac{15\pi^2\hbar^3}{(M_Pc)^2mc}n~.
\end{eqnarray}
From the above we can see that the relative correction increases only with increasing $n$ and does not depend on $m$ for the neutral bosons and increases with increasing $n$ and decreasing $m$ for the charged bosons. The relative corrections are presented in Fig(\ref{qgrfig}). We see that the QG corrections for the charged boson case require a higher $\alpha_0$ (about 5 orders of magnitude) to achieve the same magnitude of the QG correction as the neutral boson case. In other words, the corrections are much smaller for the charged boson case. This is due to the higher power of the Planck constant in the charged boson case, which significantly decreases the magnitude of the correction.

The fraction of bosons in the ground state for the neutral boson case turns out to be
\begin{eqnarray}
\label{frbr}
f_0^B=\frac{n_0}{n}=1-\left(\frac{T}{T_c}\right)^3+\alpha\frac{2\pi^4}{15\,\zeta(3)}\frac{k_B}{c}\left[\frac{T^3}{T_c^2}-\frac{T^4}{T_c^3}\right]-\beta\,150\frac{\zeta(5)}{\zeta(3)}\frac{k_B^2}{c^2}\left[\frac{T^3}{T_c}-\frac{T^5}{T_c^3}\right]~,
\end{eqnarray}
while for the charged boson case
\begin{eqnarray}
\label{frbbr}
f_0^{B\bar{B}}=\frac{n_0}{n}=1-\left(\frac{T}{T_c}\right)^2+\alpha\frac{72\zeta(3)}{\pi^2}\frac{k_B}{c}\left[\frac{T^2}{T_c}-\frac{T^3}{T_c^2}\right]-\beta\,10\pi^2\frac{k_B^2}{c^2}\left[T^2-\frac{T^4}{T_c^2}\right]~.
\end{eqnarray}
From the above we see that for $\alpha,\beta\rightarrow 0$, the standard results from Eqs.(\ref{rbfr},\ref{rbbfr}) are recovered. 
We again see that at $T=T_c$, the QG corrections vanish and $f_0=0$, as expected in standard theory, even when $\alpha,\beta\neq0$. 
Therefore, as in the non-relativistic case, the 
fraction of bosons in the ground state undergoes
QG corrections for any $T<T_c$. 

 The corrections terms from Eqs.(\ref{frbr},\ref{frbbr}) are presented in Fig.(\ref{fracr}). Since only the charged boson case is dependent on the boson species, we used helium gas to plot it. We again see that the correction has a maximum between the absolute zero and the critical temperature $T_c$ at
 \begin{eqnarray}
 T_m^B=\frac{3}{4}T_c-\frac{\beta}{\alpha}\frac{3375\zeta(5)}{\pi^4}\frac{k_B}{c}T_c^2
 \end{eqnarray}
 for the neutral case and
 \begin{eqnarray}
 T_m^{B\bar{B}}=\frac{2}{3}T_c-\frac{\beta}{\alpha}\frac{5\pi^4}{54\zeta(3)}\frac{k_B}{c}T_c^2
 \end{eqnarray}
 for the charged case. This suggests that experiments able to measure the fraction of bosons in the ground state for a relativistic BEC with high accuracy would most likely observe QG corrections near this temperature $T_m$.

\subsection{Experimental implications}

There are six observables, for which we made theoretical predictions that include QG effects. 
The observables in the non-relativistic and relativistic regimes are the three critical temperatures from Eqs.(\ref{qgtcnr},\ref{qgtcbr},\ref{qgtcbbr}) and three fractions of bosons in the ground state from Eqs.(\ref{frnr},\ref{frbr},\ref{frbbr}). 
Out of these theoretical possibilities, only the 
non-relativistic BEC can currently be realized in the lab, and 
with ever-improving measurement accuracies, we hope that some of our predicted effects may be observable in this system. We also hope that relativistic BECs can also be produced in the lab in the future, which will further open up the window for measuring potential QG effects. 

In the most optimistic scenario at present, where the detection threshold for critical temperatures is of the order $\sim10^{-10}\,\mathrm{K}$, the QG effects should be large enough to be observed for $\alpha_0\gtrsim10^{19}$ when $\beta_0=\alpha_0^2$ (the same holds when $\beta_0\sim0$) and for $\beta_0\gtrsim10^{46}$ when only quadratic QG corrections are considered (i.e., $\alpha_0=0$). In the worst case the experiments do not observe any deviations from the standard theory, but we can still constrain the QG parameters to $\alpha_0<10^{19}$ for $\beta_0=\alpha_0^2$ (or $\beta_0\sim0$) and $\beta_0<10^{46}$ for $\alpha_0=0$.

The fraction of bosons in the ground state is measured by integrating the velocity distribution in the ranges of velocities, where the gas is in the condensate state \cite{AEMWC}. The precision of such a measurement is around $10^{-2}$, i.e. about $1\%$ \cite{RCCDCW,OFM} and continually improving.
We therefore expect this precision to increase with time as well, 
and reach a stage in the foreseeable future where our predicted effects will either be measurable, or one will be able to put strict bounds on the QG parameters. The bounds on QG parameters obtained by considering the precision of measuring the fraction of bosons in the ground state are $\alpha_0<10^{25}$ for $\beta_0=\alpha_0^2$ (or $\beta_0\sim0$) and $\beta_0<10^{52}$ for $\alpha_0=0$. These bounds are not as good as the ones obtained using critical temperature and are therefore phenomenologically not yet as interesting.


\section{Conclusion}
\label{conc}

Bose-Einstein condensation 
is an interesting phenomenon, which has a variety of theoretical and experimental implications. The remnant effects of compactified dimensions and QG effects, due to GUP, on a BEC could be observable through their effects on the critical temperature and the fraction of bosons in the ground state for a BEC, with a high enough sensitivity of the experimental setup. 

The effects which arise from the presence of compactified dimensions are many orders of magnitude smaller than the current experimental capabilities, but they imply interesting bounds on the dimensions of compact spaces. If such high accuracies were to be achieved, we would get an upper and lower bound simultaneously for the radius of compact dimensions, given the topology $\mathbb{R}^d\times S^N$. This can be seen if Fig.(\ref{cdfig}), for $d=3$ and $N=1$. 
To our knowledge, this is the first time that 
such constraints on {\it both} the upper and lower bounds simultaneously has been found. Therefore this could have potentially far-reaching implications in the search for extra dimensions, which is a important ingredient in certain theories of QG such as String Theory.

By considering QG effects due to GUP, the number densities for non-relativistic and relativistic gases get modified, which we use to compute the critical temperature and fraction of bosons in the ground state. We notice that both linear and quadratic corrections increase with increasing number density $n$, for all three cases, but only the charged boson case has a dependence on mass $m$. For each of the cases, the powers of $n$, the number density of bosons, are different. 
Although a relativistic BEC has not been experimentally realized so far in the laboratory, our theoretical predictions should be useful when such a state is finally achieved. 
Therefore, as of now, the only case that can be experimentally tested is the non-relativistic one. In this, increasing the number density of a boson gas influences the magnitude of the QG corrections, and for sufficiently high densities, this increase may be by one or more orders of magnitude, 
as seen from Figs.(\ref{qgnrfig},\ref{qgrfig}).

We have obtained the QG corrections to the critical temperatures in Eqs.(\ref{qgtcnr},\ref{qgtcbr},\ref{qgtcbbr}) and fractions of particles in the ground state in Eqs.(\ref{frnr},\ref{frbr},\ref{frbbr}), by considering the linear and quadratic GUP corrections separately. If both corrections are considered simultaneously, the calculations would be much more complicated, but the results in Eqs.(\ref{qgtcnr},\ref{qgtcbr},\ref{qgtcbbr},\ref{frnr},\ref{frbr},\ref{frbbr}) would change only by a numerical factor of order $\sim\mathcal{O}(1)$ in front of the quadratic correction.

Finally, as shown in our paper, it is not necessary for the 
minimum measurable length scale that follows from QG theories to be of the order of the Planck scale. In fact, it can be a potentially
observable intermediate scale, between the electroweak and Planck scales. Our results are valid for any such intermediate scale. 
If no QG effects are observed in BEC on the other hand, we can still constrain the QG parameters to $\alpha_0<10^{19}$, assuming $\beta_0=\alpha_0^2$ (or $\beta_0\sim0$) and $\beta_0<10^{46}$, assuming $\alpha_0=0$, given the experimental accuracy. 
While this is slightly worse than the bounds that follow for example from experiments at the LHC, which gives $\alpha_0 < 10^{17}, \beta_0<10^{34}$, our bounds will continue to improve with ever increasing experimental accuracies. Furthermore, 
we do expect the predicted QG effects to be present, and detectable in the future. 
We hope to report on the further ramifications of our results elsewhere. 

\section{Acknowledgement}

We thank P. Bosso and V. Todorinov for
useful discussions. 
This work was supported by the Natural Sciences and Engineering Research Council of Canada.

\begin{appendices}

\section{}
\label{mtsm}


When we calculate predictions of any physical observable in Statistical Mechanics, we have to compute averages, because in systems with many particles we can only measure macroscopic observables of the whole system, such as temperature, pressure and volume. To compute an ensemble average of a physical, single particle quantity $Y$ over the whole energy range $\varepsilon\in[0,\infty)$, for a gas of bosons or fermions, we use the ensemble average
\begin{equation}
\label{ensavg}
    \langle Y\rangle=\int_0^\infty Y(\varepsilon)g(\varepsilon)f_{\,\!_{BE(FD)}}(\varepsilon)\mathrm{d}\varepsilon~,
\end{equation}
where
\begin{equation}
\label{dos}
    g(\varepsilon)=\frac{V(2m)^{3\slash2}\varepsilon^{1\slash2}}{4\pi^2\hbar^3}
\end{equation}
is the density of states for non-relativistic bosons,
\begin{equation}
\label{dosr}
    g(\varepsilon)=\frac{V\varepsilon\sqrt{\varepsilon^2-m^2c^4}}{2\pi^2\hbar^3c^3}
\end{equation}
is the density of states for relativistic bosons and
\begin{equation}
\label{befdd}
  f_{\,\!_{BE(FD)}}(\varepsilon)=\frac{1}{e^{\beta(\varepsilon-\mu)}\mp1}~,
\end{equation}
is the BE distribution ($-$) or FD distribution ($+$). In the above $\beta=\frac{1}{k_BT}$, $k_B$ is the Boltzmann constant, $T$ the temperature, $\varepsilon$ the energy of the particle and $\mu$ the chemical potential. 
 For any single particle quantity $Y(\varepsilon)$, all integrals given by Eq.(\ref{ensavg}), which we calculate using the BE distribution in Eq.(\ref{befdd}) (using $-$), are of the following form
 
\begin{equation}
\label{bei}
    I_\nu(\beta,\beta\mu)=\int_0^\infty\frac{\varepsilon^\nu}{e^{\beta(\varepsilon-\mu)}-1}\mathrm{d}\varepsilon=\frac{\Gamma(\nu+1)}{\beta^{\nu+1}}Li_{\nu+1}(e^{\beta\mu})~,
\end{equation}
where $\nu$ is the power of the energy in the integral, $\Gamma(\nu+1)$ is the gamma function evaluated at $\nu+1$ and 
\begin{equation}
\label{plf}
    Li_\nu(x)=\sum_{k=1}^\infty\frac{x^k}{k^\nu}
\end{equation}
is the polylogarithm function. For $x=1$, which corresponds to the case $\mu=0$, the polylogarithm function in Eq.(\ref{plf}) reduces to the well known Riemann zeta function
\begin{equation}
    \zeta(\nu)=\sum_{k=1}^\infty\frac{1}{k^\nu}~.
\end{equation}

On the other hand, all integrals, which we calculate using the FD distribution in Eq.(\ref{befdd}) (using $+$), are of the following form
\begin{equation}
\label{fdi}
    J_\nu(\beta,\beta\mu)=\int_0^\infty\frac{\varepsilon^\nu}{e^{\beta(\varepsilon-\mu)}+1}\mathrm{d}\varepsilon=-\frac{\Gamma(\nu+1)}{\beta^{\nu+1}}Li_{\nu+1}(-e^{\beta\mu})~.
\end{equation}
For $x=-1$, corresponding to $\mu=0$, the polylogarithm function in Eq.(\ref{plf}) reduces to
\begin{equation}
    Li_\nu(-1)=-\eta(\nu)~,
\end{equation}
where
\begin{equation}
    \eta(\nu)=\sum_{k=1}^\infty\frac{(-1)^{k-1}}{k^\nu}
\end{equation}
is the Dirichlet eta function. The values for the Riemann zeta and Dirichlet eta, as a function of $\nu$ (where defined) can be found numerically.


As the simplest example we can compute the number of particles in a gas of bosons, contained in a volume $V$, using the BE distribution and Eqs.(\ref{ensavg}, \ref{dos}), as
\begin{eqnarray}
\label{nbe}
    N_{\,\!_{BE}}\!\!\!\!&=&\!\!\!\!\int_0^\infty g(\varepsilon)f_{\,\!_{BE}}(\varepsilon)\mathrm{d}\varepsilon \nonumber \\
    &=&\!\!\!\!\frac{V(2m)^{3\slash2}}{4\pi^2\hbar^3}\int_0^\infty\frac{\varepsilon^{1/2}}{e^{\beta(\varepsilon-\mu)}-1}\mathrm{d}\varepsilon \nonumber \\
    &=&\!\!\!\!\frac{V(2m)^{3\slash2}}{4\pi^2\hbar^3}(k_BT)^{3\slash2}\Gamma(\tfrac{3}{2})Li_{3\slash2}(e^{\beta\mu}) \nonumber \\
    &=&\!\!\!\!\frac{V}{8\hbar^3}\left(\frac{2mk_BT}{\pi}\right)^{3/2}Li_{3\slash2}(e^{\beta\mu})~,
\end{eqnarray}
where we used Eq.(\ref{bei}) to evaluate the integral in line two. As the temperature approaches $T_c$, the chemical potential vanishes $\mu\rightarrow0$, which reduces the polylogarithm function in Eq.(\ref{nbe}) to the Riemann zeta function $\zeta(\tfrac{3}{2})\simeq2.612$. This is the regime where the Bose-Einstein condensation starts to occur.
In the same manner we can compute the number of particles in a gas of fermions, contained in a volume $V$, using the FD distribution and Eqs.(\ref{ensavg},\ref{dos}), as
\begin{eqnarray}
\label{fde}
    N_{\,\!_{FD}}\!\!\!\!&=&\!\!\!\!\int_0^\infty g(\varepsilon)f_{\,\!_{FD}}(\varepsilon)\mathrm{d}\varepsilon \nonumber \\
    &=&\!\!\!\!\frac{V(2m)^{3\slash2}}{4\pi^2\hbar^3}\int_0^\infty\frac{\varepsilon^{1/2}}{e^{\beta(\varepsilon-\mu)}+1}\mathrm{d}\varepsilon \nonumber \\
    &=&\!\!\!\!-\frac{V(2m)^{3\slash2}}{4\pi^2\hbar^3}(k_BT)^{3\slash2}\Gamma(\tfrac{3}{2})Li_{3\slash2}(-e^{\beta\mu}) \nonumber \\
    &=&\!\!\!\!-\frac{V}{8\hbar^3}\left(\frac{2mk_BT}{\pi}\right)^{3/2}Li_{3\slash2}(-e^{\beta\mu})~,
\end{eqnarray}
where we used Eq.(\ref{fdi}) to evaluate the integral in line two. The above is an exact solution for a Fermi gas at temperature $T$. In the case, when $T\longrightarrow0$, the FD distribution reduces to $f_{FD}(\varepsilon)=1$ and we would get a finite so-called Fermi energy $E_f$ as an upper limit to the integral. This would represent a degenerate Fermi gas.

\section{}
\label{lincalc}

We take a look at the operator $p_0=\sqrt{p_{0k}p_{0k}}$, where $p_{0i}=-i\hbar\frac{\partial}{\partial x_{0i}}$. Note that $p_0$ is a scalar operator. 
Explicitly it can be written as
\begin{eqnarray}
\label{p00}
p_0=\sqrt{-\hbar^2\left(\frac{\partial^2}{\partial x_{0}^2}+\frac{\partial^2}{\partial y_{0}^2}+\frac{\partial^2}{\partial z_{0}^2}\right)}=\sqrt{-\hbar^2\nabla_0^2}=\hbar\left(-\nabla_0^2\right)^{1/2}~.
\end{eqnarray}
We conveniently write it as the following, where $l$ is a non-zero constant and we have added and subtracted a $1$ inside the parenthesis. We will interpret $l$ as a length scale and therefore assume it to be positive. This also
ensures that the eigenvalues of $p_0$ are positive. 
%
%
\begin{eqnarray}
\label{p01}
p_0=\frac{\hbar}{l}\left(1-l^2\nabla_0^2-1\right)^{1/2}~.
\end{eqnarray}
We see that the above is of the form 
$(1+x)^{1/2}$, where $x=-l^2\nabla_0^2-1$, which can be 
represented as a Taylor series
\begin{eqnarray}
\label{taylor}
(1+x)^{1/2}=\sum_{m=0}^\infty c_m x^m~.
\end{eqnarray}
In the above, 
the expansion coefficients $c_m$ correspond to those 
in the Taylor series of $(1+x)^{1/2}$. However, we do not need the exact values for the remainder of our proof.  
Using the above, we can write Eq.(\ref{p01}) as
\begin{eqnarray}
\label{p02}
p_0=\frac{\hbar}{l}\sum_{m=0}^\infty c_m\left(-l^2\nabla_0^2-1\right)^m~.
\end{eqnarray}
Next, use the binomial theorem
\begin{equation}
\label{binom}
(a+b)^m=\sum_{p=0}^m {m\choose p}a^{m-p}\,b^p~,
\end{equation}
where ${m\choose p}=\frac{m!}{(m-p)!p!}$ and $a,b\in\mathbb{R}$
to rewrite Eq.(\ref{p02}) as
\begin{eqnarray}
\label{p03}
p_0=\frac{\hbar}{l}\sum_{m=0}^\infty c_m\sum_{p=0}^m{m\choose p}\left(-l^2\nabla_0^2\right)^{m-p}(-1)^p~.
\end{eqnarray}
Since the identity operator commutes with every other operator, and in particular 
$[1,\left(\nabla_0^2\right)^r]=0$, where $r\in\mathbb{N}\cup\{0\}$, 
we can rewrite Eq.(\ref{p03}) as
\begin{eqnarray}
\label{p04}
p_0=\frac{\hbar}{l}\sum_{m=0}^\infty c_m\sum_{p=0}^m{m\choose p}(-1)^p\left(l^2\right)^{m-p}\left(-\nabla_0^2\right)^{m-p}.
\end{eqnarray}
In our analysis we considered eigenfunctions of a three dimensional particle in a box $|\psi_\mathbf{n}\rangle$ with eigenvalues of operator $-\nabla_0^2$ being 
\begin{eqnarray}
-\nabla_0^2|\psi_\mathbf{n}\rangle=k_\mathbf{n}^2|\psi_\mathbf{n}\rangle~,
\end{eqnarray} 
where $k_\mathbf{n}^2=k_{nx}^2+k_{ny}^2+k_{nz}^2=\frac{\pi^2}{L^2}(n_x^2+n_y^2+n_z^2)$. Therefore, 
if we take a square of the operator $-\nabla_0^2$, we get
\begin{eqnarray}
\left(-\nabla_0^2\right)^2|\psi_\mathbf{n}\rangle\!\!\!\!&=&\!\!\!\!\left(-\nabla_0^2\right)\left(-\nabla_0^2\right)|\psi_\mathbf{n}\rangle=\left(-\nabla_0^2\right)k_\mathbf{n}^2|\psi_\mathbf{n}\rangle \nonumber \\
&=&\!\!\!\!k_\mathbf{n}^2\left(-\nabla_0^2\right)|\psi_\mathbf{n}\rangle=k_\mathbf{n}^2k_\mathbf{n}^2|\psi_\mathbf{n}\rangle=\left(k_\mathbf{n}^2\right)^2|\psi_\mathbf{n}\rangle~.
\end{eqnarray}
Similarly, for all other powers $r\in\mathbb{N}\cup\{0\}$ of the operator $-\nabla_0^2$, it can be proven by induction, that
\begin{eqnarray}
\left(-\nabla_0^2\right)^r|\psi_\mathbf{n}\rangle=\left(k_\mathbf{n}^2\right)^r|\psi_\mathbf{n}\rangle~.
\end{eqnarray}
Having all necessary information, we can now use the operator in Eq.(\ref{p04}) to compute it's eigenvalue on the eigenfunction $|\psi_\mathbf{n}\rangle$
\begin{eqnarray}
p_0|\psi_\mathbf{n}\rangle\!\!\!\!&=&\!\!\!\!\frac{\hbar}{l}\sum_{m=0}^\infty c_m\sum_{p=0}^m{m\choose p}(-1)^p\left(l^2\right)^{m-p}\left(-\nabla_0^2\right)^{m-p}|\psi_\mathbf{n}\rangle \nonumber \\
&=&\!\!\!\!\frac{\hbar}{l}\sum_{m=0}^\infty c_m\sum_{p=0}^m{m\choose p}(-1)^p\left(l^2\right)^{m-p}\left(k_\mathbf{n}^2\right)^{m-p}|\psi_\mathbf{n}\rangle \nonumber \\
&=&\!\!\!\!\frac{\hbar}{l}\sum_{m=0}^\infty c_m\sum_{p=0}^m{m\choose p}\left(l^2k_\mathbf{n}^2\right)^{m-p}(-1)^p|\psi_\mathbf{n}\rangle \nonumber \\
&=&\!\!\!\!\frac{\hbar}{l}\sum_{m=0}^\infty c_m\left(l^2k_\mathbf{n}^2-1\right)^m|\psi_\mathbf{n}\rangle \nonumber \\
&=&\!\!\!\!\frac{\hbar}{l}\left(1+l^2k_\mathbf{n}^2-1\right)^{1/2}|\psi_\mathbf{n}\rangle \nonumber \\
&=&\!\!\!\!\frac{\hbar}{l}\left(l^2k_\mathbf{n}^2\right)^{1/2}|\psi_\mathbf{n}\rangle \nonumber \\
&=&\!\!\!\!\hbar\left(k_\mathbf{n}^2\right)^{1/2}|\psi_\mathbf{n}\rangle~,
\end{eqnarray}
To compute the eigenvalue of the operator $p_0^3$, we use operators $p_0$ and $p_0^2$ consecutively on the state $|\psi_\mathbf{n}\rangle$
\begin{eqnarray}
p_0^3|\psi_\mathbf{n}\rangle\!\!\!\!&=&\!\!\!\!p_0^2p_0|\psi_\mathbf{n}\rangle=p_0^2\hbar\left(k_\mathbf{n}^2\right)^{1/2}|\psi_\mathbf{n}\rangle=\hbar\left(k_\mathbf{n}^2\right)^{1/2}p_0^2|\psi_\mathbf{n}\rangle \nonumber \\
&=&\!\!\!\!\hbar\left(k_\mathbf{n}^2\right)^{1/2}\left(-\hbar^2\nabla^2\right)|\psi_\mathbf{n}\rangle=\hbar\left(k_\mathbf{n}^2\right)^{1/2}\left(\hbar^2k_\mathbf{n}^2\right)|\psi_\mathbf{n}\rangle \nonumber \\ 
&=&\!\!\!\!\hbar^3\left(k_\mathbf{n}^2\right)^{3/2}|\psi_\mathbf{n}\rangle~.
\end{eqnarray}

To our knowledge, this is the first time that the
eigenfunctions of the $p_0^3$ operator in three spatial dimensions have been found by this method, 
thereby providing a simple solution for future research in QG phenomenology involving a linear GUP.

\section{}
\label{dp}

The QG corrected density of states is obtained in a similar way as it is obtained without QG corrections. Without QG corrections, in the continuum limit, and for the dispersion relation $\varepsilon(p)$, the number of particles, and by extension the density of states is given by
\begin{eqnarray}
\label{npar}
\sum_\mathbf{n}\approx\int\mathrm{d}^3n=\frac{V}{(2\pi\hbar)^3}\int_0^\infty\mathrm{d}^3p=\frac{V}{2\pi^2}\int_0^\infty k^2\,\mathrm{d}k=\int_0^\infty g(\varepsilon)\mathrm{d}\varepsilon~,
\end{eqnarray}
where $p=\hbar k$ and $\mathrm{d}^3p=4\pi p^2\mathrm{d}p$ were used. By using the modified dispersion relations from Eqs.(\ref{qgpib},\ref{qgpibr}), we obtain the QG corrected density of states for non-relativistic and relativistic particles respectively. We modify both $k^2$ and $\mathrm{d}k$, by expressing $k$ in terms of the particle energy $\varepsilon$ from Eq.(\ref{qgpib}) for the non-relativistic case and from Eq.(\ref{qgpibr}) for the relativistic case, in the continuum limit ($k_\mathbf{n}\longrightarrow k$ and $\varepsilon_\mathbf{n}\longrightarrow \varepsilon$). We considered the linear and quadratic GUP separately for convenience. 
Considering both contributions simultaneously would make the results change just by a numerical factor of $\mathcal{O}(1)$ in front of the quadratic term.

\subsection{Quadratic GUP}

For the quadratic QG correction ($\alpha=0$), $k$ is obtained from Eq.(\ref{qgpib}) by solving a quadratic equation for $k^2(\varepsilon)$ and the solutions are
\begin{eqnarray}
\label{k2q}
k_{1,2}^2=\left\{\begin{array}{ll}
 \displaystyle{\frac{1}{10\beta\hbar^2}\left[-1\pm\sqrt{1+40\beta m\varepsilon}\right]}~,    & \mathrm{non-relativistic} \\
  \displaystyle{\frac{1}{10\beta\hbar^2}\left[-1\pm\sqrt{1+20\beta \left(\frac{\varepsilon^2}{c^2}-m^2c^2\right)}\right]}~,   & \mathrm{relativistic}
\end{array}\right.
\end{eqnarray}
Each of the above cases gives rise to $4$ solutions. However, we restrict ourselves to $k_{1,2} \in \mathbb{R}$ {\it and} to 
$k_{1,2}>0$, it being the radius of a sphere in $k-$space. %
This reduces the number of solutions to just $1$ each. 


To obtain the QG corrected measure $\mathrm{d}k$, we calculated the derivatives of Eqs.(\ref{qgpib},\ref{qgpibr}) (for $\alpha=0$) with respect to $k$ and expressed $\mathrm{d}k$ as
\begin{eqnarray}
\label{dkq}
\mathrm{d}k=\left\{\begin{array}{ll}
  \displaystyle{\frac{\mathrm{d}\varepsilon}{\frac{\hbar^2k}{m}+\frac{10\beta\hbar^4k^3}{m}}}~,   & \mathrm{non-relativistic} \\
   & \\
  \displaystyle{\frac{\varepsilon\mathrm{d}\varepsilon}{\hbar^2c^2k+10\beta\hbar^4c^2k^3}}~,   & \mathrm{relativistic}
\end{array}\right.
\end{eqnarray}

To obtain the density of states with quadratic QG corrections, we plug the solution for $k$ from Eq.(\ref{k2q}) in Eq.(\ref{dkq}), such that the measure is now completely dependent on $\varepsilon$.
Finally, we substitute both Eqs.(\ref{k2q},\ref{dkq}) in Eq.(\ref{npar}) to obtain the QG corrected densities of states in Eqs.(\ref{qgdos}, \ref{qgdosr}) for non-relativistic and relativistic particles respectively. Note that a perturbative approach, 
dropping terms of order equal to or higher than $\mathcal{O}(\beta^2)$, was necessary to obtain the QG corrected densities of states.

\subsection{Linear GUP}

We follow a similar procedure as in the previous subsection.
For the linear QG correction ($\beta=0$), $k$ is obtained from Eqs.(\ref{qgpib},\ref{qgpibr}) by solving cubic equations for $k(\varepsilon)$, giving rise to $3$ solutions for each of the non-relativistic and relativistic cases
%
\begin{eqnarray}
\label{kl}
k=\left\{\begin{array}{ll}
 \displaystyle{\frac{1}{6\alpha\hbar}[1-\cos{(\varphi(\alpha))+\sqrt{3}\sin{(\varphi(\alpha))}}]}~,   &  \\
 & \\
 \displaystyle{\frac{1}{6\alpha\hbar}[1-\cos{(\varphi(\alpha))-\sqrt{3}\sin{(\varphi(\alpha))}}]}~,    &  \\
 & \\
 \displaystyle{\frac{1}{6\alpha\hbar}[1+2\cos{(\varphi(\alpha))}]}~,   &
\end{array}\right.
\end{eqnarray}
%
where
\begin{eqnarray}
\varphi(\alpha)=\left\{\begin{array}{ll}
  \displaystyle{\frac{1}{3}\arctan{\left(\frac{6\sqrt{6}\alpha\sqrt{m\varepsilon}\sqrt{1-54\alpha^2m\varepsilon}}{1-108\alpha^2m\varepsilon}\right)}}~,   & \mathrm{non-relativistic} \\
  & \\
  \displaystyle{\frac{1}{3}\arctan{\left(\frac{2\sqrt{27}\alpha\sqrt{\frac{\varepsilon^2}{c^2}-m^2c^2}\sqrt{1-27\alpha^2\left(\frac{\varepsilon^2}{c^2}-m^2c^2\right)}}{1-54\alpha^2\left(\frac{\varepsilon^2}{c^2}-m^2c^2\right)}\right)}}~,   & \mathrm{relativistic}
\end{array}\right.
\end{eqnarray}
Next, out the the $3$ solutions 
of Eq.(\ref{kl}), only the first is physically relevant, since the second solution is not positive and the third diverges in the limit $\alpha_0\longrightarrow 0$. We are therefore left with only $1$ solution for each case.
%
%

To obtain the QG corrected measure $\mathrm{d}k$, we calculate the derivatives of Eqs.(\ref{qgpib},\ref{qgpibr}) (for $\beta=0$) with respect to $k$ and express $\mathrm{d}k$ as
\begin{eqnarray}
\label{dkl}
\mathrm{d}k=\left\{\begin{array}{ll}
  \displaystyle{\frac{\mathrm{d}\varepsilon}{\frac{\hbar^2k}{m}-\frac{3\alpha\hbar^3k^2}{m}}}~,   & \mathrm{non-relativistic} \\
   & \\
  \displaystyle{\frac{\varepsilon\mathrm{d}\varepsilon}{\hbar^2c^2k-3\alpha\hbar^3c^2k^2}}~,   & \mathrm{relativistic}~.
\end{array}\right.
\end{eqnarray}

To obtain the density of states with linear QG corrections, we plug the solution for $k$ from Eq.(\ref{kl}) in Eq.(\ref{dkl}) for the measure to be completely dependent on $\varepsilon$. Finally, we substitute both Eqs.(\ref{kl},\ref{dkl}) in Eq.(\ref{npar}) to obtain the QG corrected densities of states in Eqs.(\ref{qgdos}, \ref{qgdosr}) for non-relativistic and relativistic particles respectively. Note that also here a perturbative approach, using Taylor series expansions, dropping terms of quadratic order or higher. 
%

\section{}
\label{RKP}

For a particle in a box without any QG corrections, we can define the `dimensionless energy' $\varepsilon^*$ as
\begin{equation}
\varepsilon^* \equiv n_x^2+n_y^2+n_z^2=\frac{2mL^2\varepsilon}{\hbar^2\pi^2}~.
\end{equation}
When we include QG corrections, we need to solve the quadratic equation, from Eq.(\ref{qgpib}) ($\alpha=0$) for $n^2 \equiv n_x^2+n_y^2+n_z^2$
\begin{eqnarray}
\label{qgsq}
n_{1,2}^2=\frac{L^2}{10\beta\pi^2\hbar^2}(-1\pm\sqrt{1+40\beta m\varepsilon_\mathbf{n}})~,
\end{eqnarray}
In what follows, we will only consider the $+$ sign, 
since the right hand side of Eq.(\ref{qgsq}) is negative (and hence $n$ imaginary, whereas 
$n_x,n_y,n_z\in\mathbb{Z}$) 
for the solution with the $-$ sign. 
%
%

Next, to get the dimensionless energy with no QG corrections for a gas of $N$ such particles, one adds up single particle energies $\varepsilon_i$ to get
\begin{equation}
\label{esum}
    \sum_{r=1}^{3N}n_r^2=\frac{2mL^2E}{\hbar^2\pi^2}
    \equiv E^*~,
\end{equation}
where $E=\varepsilon_1+\varepsilon_2+\cdots+\varepsilon_N$ and $E^*=\varepsilon^*_1+\varepsilon^*_2+\cdots+\varepsilon^*_N$. We follow the same procedure 
for the QG corrected dimensionless energy, in which using we use the Taylor expansion up to second order ($\sqrt{1+x}\simeq1+\tfrac{1}{2}x-\tfrac{1}{8}x^2$) 
in Eq.(\ref{qgsq}), and evaluate the sum in Eq.(\ref{esum}) for $N$ particles to get
\begin{eqnarray}
\label{qgde}
    \sum_{r=1}^{3N}n_r^2=\frac{2mV^{2/3}E}{\hbar^2\pi^2}-\frac{20\beta V^{2/3}m^2E_s^2}{\hbar^2\pi^2}=E^*~,
\end{eqnarray}
where $E=\varepsilon_{\mathbf{n}_1}+\varepsilon_{\mathbf{n}_1}+\cdots+\varepsilon_{\mathbf{n}_N}$,
i.e. the sum of energies of all the particles, 
$E_s^2=\varepsilon_{\mathbf{n}_1}^2+\varepsilon_{\mathbf{n}_1}^2+\cdots+\varepsilon_{\mathbf{n}_N}^2$, i.e.
the sum of its squares 
and $V^{2/3}=L^2$. 
$E_s$ is related to the total energy $E$ through $E^2=E_s^2+2E_m^2=\varepsilon_{\mathbf{n}_1}^2+\varepsilon_{\mathbf{n}_1}^2+\cdots+\varepsilon_{\mathbf{n}_N}^2+2\varepsilon_{\mathbf{n}_1}\varepsilon_{\mathbf{n}_2}+2\varepsilon_{\mathbf{n}_1}\varepsilon_{\mathbf{n}_3}+\cdots+2\varepsilon_{\mathbf{n}_2}\varepsilon_{\mathbf{n}_3}\cdots$, where $E_m^2$ is the sum of all mixed terms.

To compute the number of microstates in a 
$d$-dimensional sphere in $E^*$ space 
up to some arbitrary energy, we use $V_d(R)=\tfrac{\pi^{d/2}}{\Gamma(d/2+1)}R^d$ and we take $n^2\geq0$, so we are left with just the upper half of a sphere. Using $d=3N$ and $R=\sqrt{E^*}$, the number of microstates becomes
\begin{eqnarray}
\label{nms}
    \Sigma_N(E^*)=\left(\frac{1}{2}\right)^{3N}\left[\frac{\pi^{\tfrac{3N}{2}}}{\Gamma(\tfrac{3N}{2}+1)}\left(E^*\right)^{\tfrac{3N}{2}}\right]~,
\end{eqnarray}
in which we then plug in Eq.(\ref{qgde}) to obtain
\begin{eqnarray}
\Sigma(N,V,E)=\left(\frac{1}{2}\right)^{3N}\left[\frac{\pi^{\tfrac{3N}{2}}}{\Gamma(\tfrac{3N}{2}+1)}\left(\frac{2mV^{2/3}E}{\hbar^2\pi^2}-\frac{20\beta V^{2/3}m^2E_s^2}{\hbar^2\pi^2}\right)^{\tfrac{3N}{2}}\right]~.
\end{eqnarray}
The number of microstates in a spherical shell of thickness $\Delta$ is computed as
\begin{eqnarray}
\label{qgms}
    \Gamma(N,V,E;\Delta)\!\!\!\!&=&\!\!\!\!\frac{\partial\Sigma(N,V,E)}{\partial E}\Delta \\
    &=&\!\!\!\!\frac{\Delta}{E}\frac{\pi^{\tfrac{3N}{2}}}{(\tfrac{3N}{2}-1)!}\frac{V^N}{2^{3N}\pi^{3N}\hbar^{3N}}\left[2mE-20\beta m^2\left(E^2-2E_m^2\right)\right]^{\tfrac{3N}{2}}\frac{1-20\beta m\left(E-2E_m\tfrac{\partial E_m^{}}{\partial E}\right)}{1-10\beta m\left(E-2\tfrac{E_m^2}{E}\right)}~. \nonumber
\end{eqnarray}

To get a number of microstates in an energy shell with thickness $\Delta$, we can use the phase space integral $\Gamma(N,V,E)=\omega/\omega_0$, where $\omega_0$ is the normalization of the phase space integral, which we want to find with QG corrections and
\begin{eqnarray}
\label{psint}
    \omega\!\!\!\!&=&\!\!\!\!\int\mathrm{d}^{3N}x\int\mathrm{d}^{3N}p=V^N\int_{2m(E-\tfrac{1}{2}\Delta)\leq\sum_{i=1}^{3N}y_i^2\leq2m(E+\tfrac{1}{2}\Delta)}\cdots\int \mathrm{d}^{3N}y= \nonumber \\
    &=&\!\!\!\!V^N\frac{\Delta}{E}\frac{(2\pi mE)^{\tfrac{3N}{2}}}{(\tfrac{3N}{2}-1)!}=\Gamma(N,V,E)\omega_0~.
\end{eqnarray}
We compare Eq.(\ref{qgms}) and Eq.(\ref{psint}) to get
\begin{eqnarray}
\label{Nnorm}
\omega_0=\frac{(2\pi\hbar)^{3N}}{\left[1-10\beta m\left(E-2\tfrac{E_m^2}{E}\right)\right]^{\tfrac{3N}{2}-1}\left[1-20\beta m\left(E-2E_m\tfrac{\partial E_m^{}}{\partial E}\right)\right]}~,
\end{eqnarray}
which is valid for any arbitrary number of particles $N$. For the limit, where $N\rightarrow\infty$, Eq.(\ref{Nnorm}) reduces to $\omega_0=(2\pi\hbar)^{3N}$. For BEC we are interested in the case where $N=1$
\begin{eqnarray}
\omega_0=\frac{(2\pi\hbar)^{3}}{1-25\beta m\varepsilon}~,
\end{eqnarray}
where $\varepsilon$ is again a single particle energy.
It may be noted that 
the density of states derived using the 
modified normalization of the phase space integral
is identical to that obtained by using 
the method in Appendix \ref{dp} for quadratic GUP.

\end{appendices}

\end{document}